\documentclass[aps,showpacs,nofootinbib,superscriptaddress]{revtex4}

\usepackage{graphicx}
\usepackage{dcolumn}

\def\slashchar#1{\setbox0=\hbox{$#1$}
   \dimen0=\wd0 \setbox1=\hbox{/} \dimen1=\wd1
   \ifdim\dimen0>\dimen1 \rlap{\hbox to \dimen0{\hfil/\hfil}} #1
   \else  \rlap{\hbox to \dimen1{\hfil$#1$\hfil}} / \fi}
\def\tstrut{\vrule height2.5ex depth0pt width0pt} 
\def\jtstrut{\vrule height5ex depth0pt width0pt} 
\begin{document}
\title{  Combined nonrelativistic constituent quark model and heavy quark
effective theory study of semileptonic decays of $\Lambda_b$ and $\Xi_b$ baryons.} 
\author{C. Albertus} \affiliation{Departamento de
F\'{\i}sica Moderna, Universidad de Granada, E-18071 Granada, Spain.}
\author{ E. Hern\'andez} \affiliation{Grupo de F\'\i sica Nuclear,
Facultad de Ciencias, E-37008 Salamanca, Spain.}  \author {J. Nieves}
\affiliation{Departamento de F\'{\i}sica Moderna, Universidad de
Granada, E-18071 Granada, Spain.}
\begin{abstract} 
\rule{0ex}{3ex} 

\end{abstract}


\begin{abstract} 
\rule{0ex}{3ex} We present the results of a nonrelativistic
constituent quark model  study of the semileptonic decays
$\Lambda_b^0 \to \Lambda_c^+ l^- {\bar \nu}_l$ and $\Xi_b^0 \to
\Xi_c^+ l^- {\bar \nu}_l$ ($l=e,\mu$). We work on coordinate space, with baryon
wave functions recently obtained from a variational approach based on
heavy quark symmetry . We develop a novel expansion of the
electroweak current operator, which supplemented with heavy quark
effective theory  constraints, allows us to predict the baryon
form factors and the decay distributions for all $q^2$ (or
equivalently $w$) values accessible in the physical decays.  Our
results for the partially integrated longitudinal and transverse
decay widths, in the vicinity of the $w=1$ point, are in excellent
agreement with lattice calculations.  Comparison of our integrated
$\Lambda_b-$decay width to experiment allows us to extract the $V_{cb}$
Cabbibo-Kobayashi-Maskawa  matrix element for which we obtain a
value of $|V_{cb}| = 0.040\pm 0.005~({\rm
stat})~^{+0.001}_{-0.002}~({\rm theory})$ also in excellent agreement
with a recent determination by the DELPHI Collaboration from the
exclusive ${\bar {\rm B}^0_{\rm d}} \to {\rm D}^{*+}l^-{\bar \nu}_l$
decay. Besides for the $\Lambda_b (\Xi_b)-$decay, the longitudinal and
transverse asymmetries, and the longitudinal to transverse decay ratio
are $\langle a_L \rangle=-0.954\pm 0.001~(-0.945\pm 0.002)$  , $\langle a_T
\rangle=-0.665\pm 0.002~(-0.628\pm 0.004)$ and $R_{L/T}=1.63\pm 0.02
~(1.53\pm 0.04)$, respectively. 
\end{abstract}
\maketitle

\section{Introduction}
The understanding of the non-perturbative strong interaction effects
in the exclusive $b\to c$ semi-leptonic transition is necessary for
the determination of the $cb$ ($V_{cb}$) Cabbibo-Kobayashi-Maskawa (CKM) matrix element from the
experimentally measured rates and distributions.  A considerable
amount of work has been carried out in the meson sector, where the
ideas of heavy quark symmetry (HQS)~\cite{hqs} and heavy quark
effective theory (HQET)~\cite{hqet} were first developed. In
the theoretical side, there exist lattice
calculations~\cite{Be93}--\cite{Sh00}, and a large variety of other
theoretical analysis (HQET, dispersive bounds, quark model, sum rules,
etc.)~\cite{HQET2}--~\cite{SumRules}. From the experimental point of
view there were also an important activity and CLEO and Belle
collaborations have recent measurements of ${\rm B} \to  {\rm D}^*$
decays~\cite{Exp1}--\cite{Exp3}.

The discovery of the $\Lambda_b$ baryon at CERN~\cite{Al91}, the
discovery of most of the charmed baryons of the SU(3) multiplet on the
second level of the SU(4) lowest 20-plet~\cite{pdg04}, and the recent
measure of the semileptonic decay of the $\Lambda_b^0$~\cite{HB-exp}
make the study of the weak interactions of heavy baryons
timely. Experimental knowledge of the $\Lambda_b$ semileptonic decay
can lead to an independent estimate of $V_{cb}$ if the effects of the
strong interaction in the decay are understood. There exists an
abundant literature on the
subject~\cite{HB-Lattice}--~\cite{HB-QM}. Almost all theoretical
approaches applied to the meson sector have also been  explored for
baryons. A common drawback in most of these studies is the
impossibility of describing the decay distributions for all $q^2$
($q$ is the four momentum transferred to the leptons in the decay)
accessible values in the physical decay. Thus, lattice calculations and
HQET based approaches lead to reliable predictions in the neighborhood
of $q^2_{\rm max}= (m_{\Lambda_b}-m_{\Lambda_c})^2$, conventional sum
rule approaches are more reliable near $q^2=0$, while traditional 
nonrelativistic constituent quark models
(NRCQM's)
cannot predict differential decay rates far from $q^2_{\rm max}$.

HQS allows theoretical control of the non-perturbative aspects of the
calculation around the infinite quark mass limit. The classification
of the weak decay form factors of heavy baryons has been simplified
greatly in HQET~\cite{HQET-IW}. In addition, the $\Lambda_{Q=b,c}, \Xi_{Q=b,c}$
baryons have a particularly simple structure in that they are composed of a
heavy quark and light degrees of freedom with zero angular
momentum. At  leading order in an expansion on the heavy quark mass only one
universal form factor, the Isgur-Wise function, is required to
describe the $\Lambda_b \to \Lambda_c$ semileptonic decay. In  next to leading
order, $1/m_Q$~\cite{Lu90}, one more universal function and one mass
parameter are introduced~\cite{Ge90}. However, HQS does not determine
the universal form factors and the mass parameter, and one still needs
to employ some other non-perturbative methods.

In this work we determine the non-perturbative corrections to the
electroweak $\Lambda_b \to \Lambda_c$ matrix element by using
different NRCQM's. We use a spectator model with only one--body
current operators, and work in coordinate space, with baryon wave
functions recently obtained from a HQS based variational\footnote{In
Ref.~\cite{Al04}, we developed a rather simple method to solve the
nonrelativistic three-body problem for baryons with a heavy quark,
where we have made full use of the consequences of HQS for that
system. Thanks to HQS, the method proposed provides us with simple
wave functions, while the results obtained for the spectrum and other
observables compare quite well with the lengthly Faddeev calculations
done in~\protect\cite{si96}.}  approach~\cite{Al04}.
We propose a novel expansion of the
electroweak current operator, which allows us to predict the decay
distributions for all $q^2$ values accessible  in the physical
decay. Thus, we keep up to first order terms in the internal (small)
heavy quark momentum within the baryon, but all orders in the
transferred (large) momentum $\vec{q}$. Some preliminary results were
presented in \cite{beach04}. Now, we shall further
 impose ${\cal
O}(1/m_Q)$ accuracy HQET constraints among the form factors to improve on the
spectator model results. The paper is organized as follows. 
In Sect.~\ref{sec:cont} we introduce the form factors and their relation
to the differential decay width. Those form factors carry all 
non-perturbative QCD corrections to  the semileptonic $\Lambda_b$ and
$\Xi_b$ decays. In Sect.~\ref{sec:nrcqm}, we
relate baryon wave function with form factors, and introduce the heavy
quark internal momentum expansion (Subsect.~\ref{sec:me}). A brief
summary of the HQET predictions for these decays is outlined in
Sect.~\ref{sec:hqet}, while our results and main conclusions are
presented in Sects.~\ref{sec:res} and~\ref{sec:concl}, respectively. 
Finally, in the  Appendix some detailed formulae  can be found.

\section{Differential Decay Width and Form Factors}
\label{sec:cont}
We will focus on the $\Lambda_b (p) \to \Lambda_c (p^\prime)\, l\,
(k^\prime) {\bar \nu}_l (k)$ reaction, where $p, p^\prime, k$ and
$k^\prime$ are the four-momenta of the involved particles. The
generalization to the study of the $\Xi_b$ baryon semileptonic decay
is straightforward. In the $\Lambda_b$ rest frame, the differential
decay width reads
\begin{equation}
{\rm d}\Gamma= 
8 |V_{cb}|^2 m_{\Lambda_c} G^{\,2}  
 \frac{d^3p^\prime}{2E^\prime_{\Lambda_c} (2\pi)^3}
\frac{d^3k}{2E_{\nu_l} (2\pi)^3} \frac{d^3k^\prime}{2E^\prime_{l}
(2\pi)^3}  (2\pi)^4 \delta^4(p-p^\prime-k-k^\prime) L^{\alpha\beta} 
W_{\alpha\beta}    
\end{equation}
where\footnote{We also take $m_{\Lambda_b} = 5624$ MeV, $m_{\Xi_b}
  = 5800$ MeV and $m_{\Xi_c} = 2469$ MeV  .} $m_{\Lambda_c}
  = 2285$ MeV, and $G=1.1664\times 10^{-11}$ MeV$^{-2}$ is the Fermi
  decay constant.   $L$ and $W$ are the leptonic and hadronic
  tensors, respectively. The leptonic tensor is given by (in our
  convention, we take $\epsilon_{0123}= +1$ and the metric
  $g^{\mu\nu}=(+,-,-,-)$):
\begin{eqnarray}
L_{\mu\sigma}&=& k^\prime_\mu k_\sigma +k^\prime_\sigma k_\mu
- g_{\mu\sigma} k\cdot k^\prime + {\rm i}
\epsilon_{\mu\sigma\alpha\beta}k^{\prime\alpha}k^\beta \label{eq:lep}
\end{eqnarray}
The hadronic tensor includes all sort of non-leptonic
vertices and corresponds to the charged electroweak 
$\Lambda_b \to  \Lambda_c$ transition.  It is given by
\begin{eqnarray}
W^{\mu\sigma} &=& \frac12 \sum_{r,s}  
 \langle \Lambda_c;
\vec{p}^{\,\prime}, s|
j^\mu_{\rm cc}(0)| \Lambda_b;
\vec{p}, r   \rangle \langle \Lambda_c;
\vec{p}^{\,\prime}, s| j^\sigma_{\rm cc}(0)|  \Lambda_b;
\vec{p}, r \rangle^*
\label{eq:wmunu}
\end{eqnarray}
where $r$ and $s$ are helicity indices and  baryon states are
normalized so that $\langle \vec{p}, r | \vec{p}^{\,\prime}, s \rangle
= (2\pi)^3 (E/m)
\delta^3(\vec{p}-\vec{p}^{\,\prime})\delta_{rs}$. Finally  the
charged current is given by
\begin{equation}
j^\mu_{\rm cc} = \overline{\Psi}_c\gamma^\mu(1-\gamma_5)\Psi_b 
\end{equation}
with $\Psi_c$ and $\Psi_b$ quark fields.\\
 The non-perturbative strong
interaction effects are contained in the matrix elements of the
 weak current, $j^\mu_{\rm cc}$, which can be written in terms of six
 invariant form factors $F_i,G_i$ with $i=1,2,3$, as follows
\begin{equation}
\langle \Lambda_c; \vec{p}^{\,\prime}, s| j_\mu^{\rm cc}(0)| \Lambda_b; \vec{p}, r
\rangle = {\bar u}_{\Lambda_c}^{(s)}(\vec{p}^{\,\prime})\Big\{
\gamma_\mu\left(F_1-\gamma_5 G_1\right)+ v_\mu\left(F_2-\gamma_5
G_2\right)+v^\prime_\mu\left(F_3-\gamma_5 G_3
\right)\Big\}u_{\Lambda_b}^{(r)}(\vec{p}) \label{eq:def_ff}
\end{equation}
where $u_{\Lambda_c}$ and $u_{\Lambda_b}$ are dimensionless
  $\Lambda_c$ and $\Lambda_b$ Dirac spinors, normalized to ${\bar u} u
  = 1$, and $v_\mu = p_\mu/m_{\Lambda_b}$ ($v^\prime_\mu =
  p^\prime_\mu/m_{\Lambda_c}$) is the four velocity of the $\Lambda_b$
  ($\Lambda_c$) baryon. The form factors are functions of the velocity
  transfer $w=v\cdot v^\prime$ or equivalently of
  $q^2=(p-p^\prime)^2 = m_{\Lambda_b}^2 + m_{\Lambda_c}^2 -
  2m_{\Lambda_b}m_{\Lambda_c}w$. In the decay $\Lambda_b (p) \to
  \Lambda_c (p^\prime)\, l\, (k^\prime) {\bar \nu}_l (k)$ and for
  massless leptons, the variable $q^2$ ranges from 0 (smallest
  transfer), which corresponds to $w=w_{\rm max}=
  (m_{\Lambda_b}^2 +
  m_{\Lambda_c}^2)/2m_{\Lambda_b}m_{\Lambda_c}\approx 1.434$, to
  $q^2_{\rm max}=(m_{\Lambda_b}-m_{\Lambda_c})^2$ (highest transfer,
  final $\Lambda_c$ at rest), which corresponds to $w=1$.

The differential decay rates from transversely $(\Gamma_T)$ and
longitudinally $(\Gamma_L)$ polarized $W$'s, are given, neglecting lepton
masses,  by
(the total width is $\Gamma=\Gamma_L+\Gamma_T$)~\cite{POL}
\begin{eqnarray}
\frac{{\rm d}\Gamma_T}{{\rm d}w}&=&
\frac{G^2 |V_{cb}|^2}{12\pi^3}m_{\Lambda_c}^3\sqrt{w^2-1}\,q^2 
\Big\{ (w-1)|F_1(w)
|^2+(w+1)|G_1(w)|^2 \Big\} \nonumber\\
&&\nonumber\\
\frac{{\rm d}\Gamma_L}{{\rm d}w}&=&
\frac{G^2 |V_{cb}|^2}{24\pi^3}m_{\Lambda_c}^3\sqrt{w^2-1}
\Big\{(w-1)|{\cal F}^V(w)|^2 + (w+1)|{\cal
F}^A(w)|^2  \Big\} \nonumber\\
&&\nonumber\\
 {\cal F}^{V,A}(w) &=& \Big[ (m_{\Lambda_b}\pm m_{\Lambda_c}) F_1^{V,A} +
(1\pm w)\left(m_{\Lambda_c} F_2^{V,A}+m_{\Lambda_b}
F_3^{V,A}\right)\Big], \quad F_i^V \equiv F_i(w) , ~ F_i^A \equiv
G_i(w),~ i=1,2,3 \label{eq:dg}
\end{eqnarray}
where in the last expression the $+(-)$ sign goes together with the
$V(A)$ upper index. The polar angle distribution reads~\cite{POL}:
\begin{equation}
\frac{{\rm d}^2\Gamma}{{\rm d}w\,{\rm d}\cos\theta} = \frac38
\left(\frac{{\rm d}\Gamma_T}{{\rm d}w} +
 2\frac{{\rm d}\Gamma_L}{{\rm d}w} \right)\Big\{1+2\alpha^\prime\cos\theta +
 \alpha^{\prime\prime} \cos^2\theta \Big\}
\end{equation}
where $\theta$ is the angle between $\vec{k}^\prime$ and
$\vec{p}^{\,\prime}$ measured in the $W_{\rm off-shell}$ rest frame,
and $\alpha^\prime$ and $\alpha^{\prime\prime}$ are asymmetry parameters
which can be expressed as
\begin{eqnarray}
\alpha^\prime &=& \frac{{\rm d}\Gamma_T}{{\rm d}w} \frac{{\rm
    d}a_T}{{\rm d}w} / \Big(\frac{{\rm d}\Gamma_T}{{\rm d}w}+
    2\frac{{\rm d}\Gamma_L}{{\rm d}w}\Big), \quad \frac{{\rm
    d}a_T}{{\rm d}w} = - \frac{G^2
    |V_{cb}|^2}{6\pi^3} 
\frac{m_{\Lambda_c}^3}{\frac{{\rm d}\Gamma_T}{{\rm d}w}}
    q^2\,(w^2-1)\,F_1(w)G_1(w) \label{eq:asy1i} \\
&&\nonumber\\
\alpha^{\prime\prime} &=& \Big (\frac{{\rm d}\Gamma_T}{{\rm d}w}  
   - 2\frac{{\rm d}\Gamma_L}{{\rm d}w}\Big)
/ \Big(\frac{{\rm d}\Gamma_T}{{\rm d}w}+
    2\frac{{\rm d}\Gamma_L}{{\rm d}w}\Big) \label{eq:asy1f}
\end{eqnarray} 
There are other asymmetry parameters if the successive hadronic cascade
decay $\Lambda_c \to a+b$, where $a$ ($J_a=1/2$) and $b$ ($J_b=0$) are
hadrons, is considered. Two new angles are usually defined,
$\Theta_\Lambda$ the angle between the $\Lambda_c$ momentum in
the $\Lambda_b$ rest frame and the $a$ hadron momentum in the $\Lambda_c$ 
rest frame, 
and $\chi$  the relative azimuthal angle between the decay planes
defined by the three-momenta of the $l$, $\nu$ leptons and the three-momenta of
the $a,b$ hadrons. The decay 
distributions
with respect to these two angles read~\cite{POL}:
\begin{equation}
\frac{{\rm d}^2\Gamma}{{\rm d}w\,{\rm d}\cos\theta_\Lambda} \propto 1 +
P_L \alpha_\Lambda \cos \theta_\Lambda, \qquad 
\frac{{\rm d}^2\Gamma}{{\rm d}w\,{\rm d}\chi} \propto 1 -
\frac{3\pi^2}{32\sqrt 2}\gamma \alpha_\Lambda \cos \chi
\end{equation}
where $\alpha_\Lambda$ is the asymmetry parameter in the $\Lambda_c$
hadronic decay (for the non-leptonic decays $\Lambda_c \to
\Lambda \pi$ and $\Lambda_c \to \Sigma \pi$ one has:
$\alpha_{\Lambda_c^+ \to \Lambda \pi^+}=-0.94^{+0.24}_{-0.08}$~\cite{CLEO93},
  $-0.96 \pm 0.42$\cite{ARGUS} and
   $\alpha_{\Lambda_c^+ \to \Sigma^+ \pi^0}=-0.45 \pm 0.32$~\cite{CLEO93}), and $P_L$ (longitudinal polarization of
the daughter baryon $\Lambda_c$) and $\gamma$ are given by
\begin{eqnarray}
P_L &=& \Big (\frac{{\rm d}\Gamma_T}{{\rm d}w} \frac{{\rm
    d}a_T}{{\rm d}w} + \frac{{\rm d}\Gamma_L}{{\rm d}w} \frac{{\rm
    d}a_L}{{\rm d}w} \Big ) / \Big(\frac{{\rm d}\Gamma_T}{{\rm d}w}+
    \frac{{\rm d}\Gamma_L}{{\rm d}w}\Big), \quad \frac{{\rm
    d}a_L}{{\rm d}w} = - \frac{G^2
    |V_{cb}|^2}{12\pi^3} 
\frac{m_{\Lambda_c}^3}{\frac{{\rm d}\Gamma_L}{{\rm d}w}}
   (w^2-1)\,{\cal F}^V(w){\cal F}^A(w) \label{eq:asy2i} \\
&&\nonumber\\
\gamma &=& \Big ( \frac{G^2
    |V_{cb}|^2}{6\sqrt 2\pi^3} m_{\Lambda_c}^3\,\sqrt{q^2}
   \,\sqrt{w^2-1}\,\left\{(w+1){\cal F}^A(w)G_1(w)-(w-1){\cal
    F}^V(w)F_1(w)\right\}\Big) / \Big(\frac{{\rm d}\Gamma_T}{{\rm d}w}+
    \frac{{\rm d}\Gamma_L}{{\rm d}w}\Big) \label{eq:asy2f}
\end{eqnarray} 
The asymmetry parameters introduced in
Eqs.~(\ref{eq:asy1i}-\ref{eq:asy1f}) and
Eqs.~(\ref{eq:asy2i}-\ref{eq:asy2f}) are functions of the velocity
transfer $w$. On averaging over $w$, the numerators and denominators
are integrated separately and  thus we have
\begin{eqnarray}
\langle a_T \rangle &=& - \frac{G^2 |V_{cb}|^2}{6\pi^3}
\frac{m_{\Lambda_c}^3}{\Gamma_T} \int_0^{w_{\rm max}}
q^2\,(w^2-1)\,F_1(w)G_1(w) {\rm d }w \label{eq:at}\\
 &&\nonumber\\
\langle a_L \rangle &=& - \frac{G^2 |V_{cb}|^2}{12\pi^3}
\frac{m_{\Lambda_c}^3}{\Gamma_L} \int_0^{w_{\rm max}}
(w^2-1)\,{\cal F}^V(w){\cal F}^A(w) {\rm d }w \\
&&\nonumber\\
\langle \gamma \rangle &=& \frac{G^2
    |V_{cb}|^2}{6\sqrt 2\pi^3} \frac{m_{\Lambda_c}^3}{\Gamma} 
   \int_0^{w_{\rm max}}\sqrt{q^2}
   (w^2-1)^{\frac12}\,\left\{(w+1){\cal F}^A(w)G_1(w)-(w-1){\cal
    F}^V(w)F_1(w)\right\}{\rm d }w \\
 &&\nonumber\\
\langle \alpha^\prime \rangle &=& \frac{\langle a_T
  \rangle}{1+2R_{L/T}}, \quad \langle \alpha^{\prime\prime} \rangle = 
\frac{1-2R_{L/T}
 } {1+2R_{L/T}}, \quad \langle P_L \rangle = \frac{\langle a_T \rangle + R_{L/T}\langle a_L
  \rangle }{1+R_{L/T}}  \qquad R_{L/T} = \frac{\Gamma_L}{\Gamma_T} \label{eq:rlt} 
\end{eqnarray}
\section{Baryon Wave Functions and Form Factors}
\label{sec:nrcqm}

\begin{figure}[t]
\vspace{-3cm}
\centerline{\includegraphics[height=25cm]{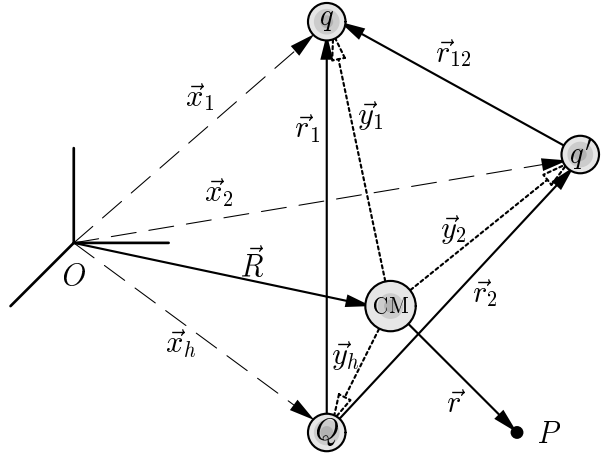}}
\vspace{-15cm}
\caption{\footnotesize Definition of different coordinates used through this
work.}\label{fig:coor}
\end{figure}

Baryon wave functions are taken from our previous work in
Ref.~\cite{Al04}, where different non-relativistic Hamiltonians ($H$)
for the three quark ($q,q^\prime, Q$, with\footnote{$l$ denotes a
light quark of flavor $u$ or $d$} $q, q^\prime =l$ or $s$ and $Q=c$ or
$b$) system of the type
\begin{eqnarray}
H&=&  \sum_{i=q,q^\prime,Q} \left (m_i
-\frac{\vec{\nabla}_{x_i}^2}{2m_i}\right ) +
V_{qq^\prime} + V_{Qq}+V_{Q q^\prime }
\end{eqnarray}
were used. In the above equation $m_q,m_{q^\prime}$ and $m_Q$
are constituent quark masses, and the quark-quark interaction terms, $V_{ij}$,
depend on the quark spin-flavor quantum numbers and the quark
coordinates ($\vec{x}_1, \vec{x}_2$ and $\vec{x}_h$ for the
$q,q^\prime$ and $Q$ quarks respectively, see
Fig.~\ref{fig:coor}). 

\subsection{Intrinsic Hamiltonian}

We briefly outline here the procedure followed in \cite{Al04}. To
separate the Center of Mass (CM) free motion, we went to the heavy
quark frame ($\vec{R},\vec{r}_1,\vec{r}_2$), where $\vec{R}$ and
$\vec{r}_1$ ($\vec{r}_2$) are the CM position in the LAB frame and the
relative position of the $q$ ($q^\prime$) quark  with respect to the
heavy $Q$ quark.  In this frame, the Hamiltonian reads
\begin{eqnarray}
H&=&
-\frac{\vec\nabla_{\vec{R}}^2}{2 M_{\rm tot}} +
H^{\rm int} \label{eq:cm}\\ H^{\rm
int}&=&\sum_{i=q,q^\prime} h^{sp}_i +
V_{qq^\prime}(\vec{r}_1-\vec{r}_2,spin) -
\frac{\vec\nabla_1\cdot\vec\nabla_2}{m_Q} + \sum_{i=q,q^\prime,Q} m_i
\label{eq:hintr} \\
h^{sp}_i &=& -\frac{\vec\nabla_i^2}{2\mu_i} + V_{Qi}(\vec
r_i,spin), \quad i=q,q^\prime  \label{eq:defhsp}
\end{eqnarray}
where $M_{\rm tot}$ is the sum of quark masses, $
\left(m_q+m_{q^\prime}+m_Q\right)$, $\mu_{q,q'} = \left (
1/m_{q,q^\prime} + 1/m_Q\right)^{-1}$ and $\vec\nabla_{1,2} =
\partial/\partial_{\vec{r}_1,\vec{r}_2}$. The intrinsic Hamiltonian
$H^{\rm int}$ describes the dynamics of the baryon and we used a
variational approach to solve it~\cite{Al02}.  $H^{\rm int}$ consists
of the sum of two single particle Hamiltonians ($h^{sp}_i$), which
describe the dynamics of the light quarks in the mean field created by
the heavy quark, plus the light--light interaction term, which
includes the Hughes-Eckart term ($\vec \nabla_1 \cdot \vec\nabla_2
$). In Ref.~\cite{Al04}, several quark-quark interactions, fitted to
the meson spectra, were used to predict charmed and bottom baryon
masses and some static electromagnetic properties. Furthers details
can be found there. 

\subsection{$\Lambda_{b,c}$ and $\Xi_{b,c}$ Wave Functions and HQS}

To solve the intrinsic Hamiltonian of Eq.~(\ref{eq:hintr}), a HQS
inspired variational approach was used in Ref.~\cite{Al04}.  HQS is an
approximate SU($N_F$) symmetry of QCD, being $N_F$ the number of heavy
flavors. This symmetry appears in systems containing heavy quarks
with masses much larger than any other energy scale ($\eta$ =
$\Lambda_{QCD}$, $m_u$, $m_d$, $m_s$,\ldots) controlling the dynamics
of the remaining degrees of freedom.  For baryons containing a heavy
quark, and up to corrections of the order ${\cal O}(\frac{\eta}{m_Q})$,
HQS guarantees that the heavy baryon light degrees of freedom quantum
numbers (spin, orbital angular momentum and parity) are always well
defined. We took advantage of this fact in Ref.~\cite{Al04} in
choosing the family of variational wave functions. Assuming that the ground
states of the baryons are in s--wave and a complete symmetry of the
wave function under the exchange of the two light quarks ($u,d,s$)
flavor, spin and space degrees of freedom (SU(3) quark model), the
wave functions read ($I$, and $S_{\rm light}^\pi$ are the isospin, and
the spin parity of the light degrees of freedom)\footnote{An obvious
notation has been used for the isospin--flavor ($|I,M_I\rangle _I$,
$|ls\rangle $ or $|sl\rangle $) and spin ($|S,M_S\rangle _{S_{\rm
light}}$) wave functions of the light degrees of freedom.}
\begin{itemize}
\item {\it $\Lambda-$type baryons: $I=0,~S_{\rm light}^\pi=0^+$}
\begin{eqnarray}
|\Lambda_Q; J=\frac12, M_J\left. \right \rangle  &=& \Big \{  |00 \rangle _I 
\otimes  |0 0 \rangle _{S_{\rm light}} \Big \} 
\Psi_{ll}^{\Lambda_Q} (r_1,r_2,r_{12}) \otimes  |Q; M_J\rangle  \label{eq:sim}
\end{eqnarray}
where the spatial wave function, since we are assuming $s-$wave
baryons, can only depend on the relative distances $r_1$, $r_2$ and
$r_{12}=|\vec{r}_1-\vec{r}_2|$.  In addition $\Psi_{ll}^{\Lambda_Q}
(r_1,r_2,r_{12}) = \Psi_{ll}^{\Lambda_Q} (r_2,r_1,r_{12})$ to
guarantee a complete symmetry of the wave function under the exchange
of the two light quarks ($u,d$) flavor, spin and space degrees of
freedom. Finally $M_J$ is the baryon total angular momentum third
component\footnote{Note, that SU(3) flavor symmetry (SU(2), in the
case of the $\Lambda_Q$ baryon) would also allow for a component in
the wave function of the type
\begin{equation}
 \sum_{M_SM_Q} (\frac12 1 \frac12 | M_QM_SM_J)\Big \{ |00 \rangle _I
\otimes |1 M_S \rangle _{S_{\rm light}} \Big \}
\Theta_{ll}^{\Lambda_Q} (r_1,r_2,r_{12}) \otimes |Q; M_Q\rangle
\label{eq:antis}
\end{equation}
with $\Theta_{ll}^{\Lambda_Q} (r_1,r_2,r_{12}) =
-\Theta_{ll}^{\Lambda_Q} (r_2,r_1,r_{12})$ (for instance terms of the
type $r_1-r_2$), and where  the real numbers $(j_1j_2j|m_1m_2m)=\langle
j_1m_1j_2m_2|jm\rangle $ are Clebsh-Gordan coefficients. This
component is forbidden by HQS in the limit $m_Q \to \infty$, where
$S_{\rm light}$ turns out to be well defined and set to zero for
$\Lambda_Q-$type baryons. The most general SU(2) $\Lambda_Q$ wave
function will involve a linear combination of the two components,
given in Eqs.~(\ref{eq:sim}) and (\ref{eq:antis}). Neglecting ${\cal
  O}(\eta/m_Q)$ corrections, HQS imposes an additional constraint,
which justifies the use of a wave function of the type of that given
in Eq.~(\ref{eq:sim}) with the obvious simplification of the three
body problem. Within a spectator model for the $\Lambda_b-$decay,
in which the light degrees of freedom remain unaltered, and due to the
orthogonality in the spin space, taking into account the $S_{\rm
  light}=1$ components of the $\Lambda_Q$ wave functions would lead to
${\cal O}(\eta^2/m_Q^2)$ corrections to the transition form factors of
Eq.~(\protect\ref{eq:def_ff}).}.

\item {\it $\Xi-$type baryons: $I=\frac12, S_{\rm light}^\pi=0^+$}
\begin{eqnarray}
|\Xi_Q; J=\frac12, M_J; M_T\left. \right \rangle  &=& \frac{1}{\sqrt2}\Big\{ 
|ls\rangle \Psi_{ls}^{\Xi_Q}(r_1,r_2,r_{12}) -
|sl\rangle \Psi_{sl}^{\Xi_Q}(r_1,r_2,r_{12}) \Big \} \otimes |00\rangle _{S_{\rm
light}} 
\otimes |Q; M_J\rangle 
\end{eqnarray}
where the isospin third component of the baryon, $M_T$, is that of the light
quark $l$ ($1/2$ or $-1/2$ for the $u$ or the $d$ quark, respectively). 
\end{itemize}
The spatial wave function\footnote{Its normalization is given by
\begin{equation}1=\int d^3r_1 \int d^3 r_2 \left |\Psi_{qq^\prime}^{B_Q}(r_1,r_2,r_{12})
\right |^2 = 8\pi^2 \int_0^{+\infty}dr_1~r_1^2~
\int_0^{+\infty}dr_2~r_2^2 \int_{-1}^{+1} d\mu~\left
|\Psi_{qq^\prime}^{B_Q}(r_1,r_2,r_{12}) \right |^2 \end{equation}
where $\mu$ is the cosine of the angle formed by $\vec{r}_1$ and $\vec{r}_2$.},
$\Psi_{qq^\prime}^{B_Q}(r_1,r_2,r_{12})$, was determined in~\cite{Al04}
by use of the variational principle $\delta \langle B_Q | H^{\rm int}| B_Q
\rangle = 0$, and can be easily reconstructed from Tables X and XI of
that reference.
\subsection{The $\langle \Lambda_c; \vec{p}^{\,\prime},s|
j_\mu^{\rm cc}(0)| \Lambda_b; \vec{p}, r \rangle$ and $\langle \Xi_c;
 \vec{p}^{\,\prime}, s| j_\mu^{\rm cc}(0)| \Xi_b; \vec{p}, r \rangle$ 
Matrix Elements}
\label{sec:qm-f}

 We will first focus on the $\Lambda_b \to \Lambda_c $ matrix element.
 Within a NRCQM and considering only one--body current operators
 (spectator approximation) we have in the $\Lambda_b$ rest frame
\begin{eqnarray}
\left<\Lambda_c; \vec{p}\,',s| 
j_{cc}^\alpha(0) |\Lambda_b;
\vec{0},r \right>&=& \sqrt{\frac{E_{\Lambda_c}^\prime}{m_{\Lambda_c}}}
 \int d^3q_1d^3q_2 d^3q_h d^3q_h^\prime\ 
\sqrt{\frac{m_b}{E_b(\vec{q}_h)}}\sqrt{\frac{m_c}{E_c(\vec{q}_h^{\,\prime})}}
\Big[\bar{u}_c^{(s)}(\vec{q}^{\,\prime}_h)\gamma^\alpha(1-\gamma_5)u_b^{(r)}
(\vec{q}_h)\Big] \nonumber\\
&\times& [\phi^{\Lambda_c}_{\vec{p}^{\,\prime}}(\vec{q}_1,\vec{q}_2,
\vec{q}^{\,\prime}_h)]^*\, \phi^{\Lambda_b}_{\vec{0}}
(\vec{q}_1,\vec{q}_2,\vec{q}_h) \label{eq:me_mom}
\end{eqnarray}
with $\vec{p}^{\,\prime}=\vec{p}-\vec{q}=-\vec{q}$, and $u_c$ and $u_b$ charm
and bottom quark Dirac spinors. The wave functions in momentum
space appearing in the above equation are the Fourier transformed of
those in coordinate space
\begin{equation}
\phi^{\Lambda_Q}_{\vec{P}}(\vec q_1,\vec q_2,\vec q_h)=
\int \frac{d^3x_1}{(2\pi)^\frac{3}{2}}
\frac{d^3x_2}{(2\pi)^\frac{3}{2}} \frac{d^3x_h}{(2\pi)^\frac{3}{2}} 
 e^{-{\rm i}(\vec{q}_1\cdot\vec{x}_1+\vec{q}_2\cdot\vec{x}_2
+\vec{q}_h\cdot\vec{x}_h)}\
\psi_{\vec{P}}^{\Lambda_Q}(\vec{x}_1,\vec{x}_2,\vec{x}_h)
\end{equation}
where the  spatial wave function of the $\Lambda_Q$ baryon with total
momentum $\vec{P}$ (see Eq.~(\ref{eq:cm})) is given by
\begin{equation}
\psi_{\vec{P}}^{\Lambda_Q}(\vec{x}_1,\vec{x}_2,\vec{x}_h)=
\frac{e^{i\vec{P}\cdot\vec{R}}}{(2\pi)^{\frac{3}{2}}}
\Psi^{\Lambda_Q}_{ll}(r_1,r_2,r_{12})
\end{equation}
with $\Psi^{\Lambda_Q}_{ll}(r_1,r_2,r_{12})$  defined in the previous
subsection.  The actual calculations are done in coordinate space, and
we find
\begin{eqnarray}
\left<\Lambda_c; -\vec{q},s| j_{cc}^\alpha(0) |\Lambda_b; \vec{0},r
\right>&=& \sqrt{\frac{E_{\Lambda_c}^\prime}{m_{\Lambda_c}}} \int
d^3r_1 d^3r_2 \, e^{{\rm i}\vec{q}\cdot\left(m_q\vec{r}_1+
m_{q'}\vec{r}_2 \right)/M_{\rm tot}^c}
[\Psi^{\Lambda_c}_{ll}(r_1,r_2,r_{12})]^*\nonumber\\
&\times&\Big\{\sqrt{\frac{m_b}
{E_b(\vec{l}\ )}}\sqrt{\frac{m_c}{E_c(\vec{l}\,')}}
\Big[\bar{u}_c^{(s)}(\vec{l}^{\,\prime})\gamma^\alpha(1-\gamma_5)u_b^{(r)}
(\vec{l}\ )\Big] \Big\} \Psi^{\Lambda_b}_{ll}(r_1,r_2,r_{12}) \label{eq:me}
\end{eqnarray}
with the operators $\vec{l}={\rm i}\vec{\nabla}_{\vec{r}_1}+ {\rm
i}\vec{\nabla}_{\vec{r}_2}$ and $\vec{l}^\prime=\vec{l}-\vec{q}$
acting on the $\Lambda_b$ intrinsic wave function. Finally, the flavor
of the light quarks ($q,q'$) are {\it up} and {\it down} and $M^c_{\rm
tot}=m_u+m_d+m_c$, with $m_u=m_d$ as dictated by SU(2)--isospin
symmetry. 

The $\Xi_b \to \Xi_c $ matrix element is easily obtained from the
results above, by using $\Psi_{ls}^{\Xi_Q}$ and $m_s$ instead of 
$\Psi_{ll}^{\Lambda_Q}$ and $m_{q^\prime}=m_u=m_d$, respectively.

\subsection{Heavy Quark Internal Momentum Expansion and Form Factor Equations}
\label{sec:me}
Taking $\vec{q}$ in the positive $z$ direction and by comparing both
sides of Eq~(\ref{eq:me}) for the spin flip $\alpha= 1~{\rm or}~2 $
and spin non-flip $\alpha=0$ and $\alpha=3$ components, all form
factors $F's$ and $G's$ can be found. The main problem lies on the
operatorial nature of the right hand side of Eq.~(\ref{eq:me}), which
requires of some approximations to make its evaluation feasible. Non
relativistic expansions of the involved momenta in Eq.~(\ref{eq:me})
are usually performed~\cite{HB-NRQM}, but  this is
only justified near $q^2_{\rm max}$.
\begin{table}[t]
\begin{center}
\begin{tabular}{cr|rcl}\hline
&&&&\\
\multicolumn{2}{c|}{Vector}  & & & \\
$\alpha=0$, & spin non--flip \,  & \,
  $\hat{F}_1+\hat{F}_2+\frac{E^\prime_{\Lambda_c}}{m_{\Lambda_c}}\hat{F}_3$&=& 
${\cal I}+  \frac{\vec{q}^{\,2}{\cal K}}{2(E_c+m_c)}\left(\frac{m_c}
{E^2_c}-\frac{1}{m_b}\right)$ \\\jtstrut
$\alpha=3$, & spin non--flip \, & \,
  $\frac{|\vec{q}\,|}{E^\prime_{\Lambda_c}+m_{\Lambda_c}}\hat{F}_1
 + \frac{|\vec{q}\,|}{m_{\Lambda_c}}\hat{F}_3$&=& $
  \frac{|\vec{q}\,|{\cal I}}{E_c+m_c} - \frac{|\vec{q}\,|{\cal K}}{2}
  \Big( \frac{m_c}{E_c^2} + \frac{1}{m_b} \Big)$\\\jtstrut
$\alpha=2$, & spin flip \, & \,
  $\frac{|\vec{q}\,|}{E^\prime_{\Lambda_c}+m_{\Lambda_c}}\hat{F}_1$ &=& $
  \frac{|\vec{q}\,|{\cal I}}{E_c+m_c} - \frac{|\vec{q}\,|{\cal K}}{2}
  \Big( \frac{m_c}{E_c^2} - \frac{1}{m_b} \Big) $
  \\
&&&&\\\hline
&&&&\\
\multicolumn{2}{c|}{Axial}  & & & \\
$\alpha=0$, & spin non--flip\,  & \,
  $\frac{|\vec{q}\,|}{E^\prime_{\Lambda_c}+m_{\Lambda_c}}
\left (-\hat{G}_1+\hat{G}_2
+\frac{E^\prime_{\Lambda_c}}{m_{\Lambda_c}}\hat{G}_3\right)$&=&$
-\frac{|\vec{q}\,|{\cal I}}{E_c+m_c} +
\frac{|\vec{q}\,|{\cal K}}{2}\Big( \frac{m_c}{E_c^2} + \frac{1}{m_b}
\Big)    
$ \\\jtstrut
$\alpha=3$, & spin non--flip \, &
\,  $\hat{G}_1 - 
\frac{\vec{q}^{\,2}}{m_{\Lambda_c}\left(E^\prime_{\Lambda_c}+m_{\Lambda_c}
\right)}\hat{G}_3$&=& 
${\cal I}+ \frac{\vec{q}^{\,2}{\cal K}}{2(E_c+m_c)}\left(\frac{m_c}
{E^2_c}-\frac{1}{m_b}\right)$
\\\jtstrut
$\alpha=1$, & spin flip \,  &
 \, $\hat{G}_1$ &=& ${\cal I}+ \frac{\vec{q}^{\,2}{\cal K}}
{2(E_c+m_c)}\left(\frac{m_c}
{E^2_c}+\frac{1}{m_b}\right)$ \\
&&&&\\\hline
\end{tabular}
\end{center}
\caption{\footnotesize Equations used to determine the $\Lambda_b \to \Lambda_c$
  transition form factors. The {\it hat} form factors and baryon
  integrals (${\cal I}$ and ${\cal K}$ ) are given in
  Eqs.~(\protect\ref{eq:hats})--(\protect\ref{eq:j}).}
\label{tab:eqs}
\end{table}
With the $\Lambda_b$ baryon at rest, $\vec{l}$ in Eq.~(\ref{eq:me}) is
an internal momentum which is much smaller than any of the heavy quark
masses. On the other hand, the transferred momentum $\vec{q}$, which
coincides, up to a sign, with the total momentum carried out by the
$\Lambda_c$ baryon, can be large (note that $|\vec{q}\,|=m_{\Lambda_c}
\sqrt{w^2-1}$ and at $q^2=0$, $|\vec{q}\,(w=w_{\rm max})| \approx
m_{\Lambda_b}/2$). We have expanded the right hand side of
Eq.~(\ref{eq:me}), neglecting second order terms in $\vec{l}$, but
keeping all orders in $\vec{q}$. For instance, this expansion for the
charm quark energy gives: $E_c(\vec{l}\,') \approx
E_c(\vec{q}\,)(1-\vec{l}\cdot\vec{q}/E_c^2(\vec{q}\,))+ {\cal
O}(\vec{l}^2/m_Q^2)$, with $E_c(\vec{q}\,) \equiv E_c =
(m_c^2+\vec{q}^{\,2})^{1/2}$.  Thanks to this novel expansion
of the electroweak current
operator, in
which $\vec{q}$ is exactly treated,  we are able to predict the decay distributions for all $q^2$
values accessible in the physical decays, improving in this manner on
the existing NRCQM calculations. Finally, we get the form factors from
two (vector and axial) subsets of three equations with three unknowns
($F's$ and $G's$).  For the $\Lambda_b \to \Lambda_c$ transition, these
equations are compiled in Table~\ref{tab:eqs}.  The {\it hat} form factors and
the dimensionless baryon integrals (${\cal I}$ and ${\cal K}$ )
appearing in the table are given by
\begin{eqnarray}
{\hat F}_i (w) &=& \left (
\frac{E^\prime_{\Lambda_c}+m_{\Lambda_c}}{2E^\prime_{\Lambda_c}}\right)^\frac12
\left (\frac{2E_c}{E_c + m_c}\right)^\frac12 F_i (w), 
~ {\hat G}_i (w) = \left (
\frac{E^\prime_{\Lambda_c}+m_{\Lambda_c}}{2E^\prime_{\Lambda_c}}\right)^\frac12
\left (\frac{2E_c}{E_c + m_c}\right)^\frac12 G_i (w), \quad i = 1,2,3 \label{eq:hats}\\ 
{\cal I}(w)&=&  
\int d^3r_1 d^3r_2 \, e^{{\rm i}\vec{q}\cdot\left(m_q\vec{r}_1+
m_{q'}\vec{r}_2 \right)/M_{\rm tot}^c}
[\Psi^{\Lambda_c}_{ll}(r_1,r_2,r_{12})]^*
\Psi^{\Lambda_b}_{ll}(r_1,r_2,r_{12})\label{eq:i}\\ 
{\cal K}(w)&=& \frac{1}{\vec{q}^{\,\,2}}
\int d^3r_1 d^3r_2 \, e^{{\rm i}\vec{q}\cdot\left(m_q\vec{r}_1+
m_{q'}\vec{r}_2 \right)/M_{\rm tot}^c}
[\Psi^{\Lambda_c}_{ll}(r_1,r_2,r_{12})]^* [\vec{l}\cdot\vec{q}\,] 
\Psi^{\Lambda_b}_{ll}(r_1,r_2,r_{12}) \label{eq:j}
\end{eqnarray}
For degenerate transitions ($m_b=m_c=m_Q$), the baryon factors ${\cal
I}(w)$ and ${\cal K}(w)$ are related, ie $2{\cal K}(w)/{\cal I}(w) =
(m_q+m_{q^\prime}) /(m_q+m_{q^\prime}+m_Q)$, as can be deduced from a
integration by parts in Eq.~(\ref{eq:j}). By means of a partial wave
expansion and after a little of Racah algebra, the integrals get
substantially simplified. Explicit expressions can be found in the
Appendix.

Baryon number conservation implies that $F(1)=\sum_i F_i (1)=1$ in the
limit of equal baryon states. The first equation of
Table~\ref{tab:eqs} leads to $\sum_i F_i (1) = {\cal I} (1) $, since
$w=1$ implies $|\vec{q}\,|=0$. Besides, ${\cal I}(1)$ accounts for the
overlap between the charmed and bottom baryon wave functions and
therefore it takes the value 1 for equal baryon states, accomplishing
exact  baryon number conservation. In general, vector current
conservation for degenerate transitions imposes the restriction
$F_2(w)=F_3(w)$, which is violated within the spectator approximation
assumed in this work. Thus for instance at zero recoil, we find
$F_2(1)-F_3(1)= 1-m_{\Lambda_Q}/M_{\rm tot}$, and thus we do not get
vector current conservation because of baryon binding terms. Two body
currents induced by inter--quark interactions are needed to conserve the
vector current.

The corresponding $\Xi_b$ decay quantities are obtained from the above
expressions by means of the substitutions mentioned at the end of
Subsect.~\ref{sec:qm-f}.  Note that, ${\cal I}$ and ${\cal K}$ depend
on both the heavy and light flavors, hence, and for the sake of clarity,
from now on we will use the notation ${\cal I}^{cb}_{\Lambda}$ or
${\cal I}^{cb}_{\Xi}$ for the $\Lambda_b$ and $\Xi_b$ decays
, and a similar notation for the ${\cal K}$ factors.

\section{HQET and Form Factors}
\label{sec:hqet}

When all energy scales relevant in the problem are much smaller than
the heavy quark masses, HQS is an excellent tool to understand charm
and bottom physics.  Close to zero
recoil ($w=1$) and at  leading order in the heavy quark mass expansion, only one
universal (independent of the heavy flavors) form factor, the
Isgur-Wise function\footnote{Note that, though called in the same manner,
because of the different light cloud, this function is different to
that entering in the study of ${\rm B}\to {\rm D}$ and ${\rm B}\to
{\rm D}^*$ semileptonic
transitions.}  ($\xi^{\rm ren}$) is required to describe the
$\Lambda_b \to \Lambda_c$ semileptonic decay. To  next order,
$1/m_Q$, one more universal ($\chi^{\rm ren}$) function and one mass
parameter (${\bar\Lambda}$) are introduced.  These functions, and also
the form--factors, depend on the heavy baryon light cloud flavor, and
thus in general they will be different for $\Xi-$ transitions, though
one expect small deviations thanks to the SU(3)-flavor symmetry.

We compile here some useful results from Ref.~\cite{HQET2,Neubert94}, where 
more details can be found. Including $1/m_Q$ corrections the $\Lambda_b
\to  \Lambda_c$ form factors factorize in the form
\begin{eqnarray}
F_i(w) &=& N_i(w) \hat{\xi}_{cb}(w) + {\cal O}(1/m_Q^2),  \quad G_i(w) =
N_i^5(w) \hat{\xi}_{cb}(w) + {\cal O}(1/m_Q^2), \qquad i=1,2,3 
\label{eq:deff}
\end{eqnarray}
 \begin{eqnarray}
\hat{\xi}_{cb}(w)&=&  \xi^{\rm ren}(w) + \Big(\frac{\bar \Lambda}{2m_b} +
\frac{\bar \Lambda}{2m_c} \Big)\left[2 \chi^{\rm ren}(w) +
  \frac{w-1}{w+1}\xi^{\rm ren}(w)  \right] \label{eq:defxi}
\end{eqnarray}
where the coefficients $N_i, N_i^5$ contain both radiative
($\hat{C}_i,\hat{C}^5_i$)\footnote{They are known up to order
$\alpha_s^2 (z{\rm ln}z)^n$, where $z=m_c/m_b$ is the ratio of the
heavy--quark masses and $n=0,1,2$} and $1/m_Q$ corrections. ${\bar
\Lambda}$ is the binding energy of the heavy quark in the
corresponding $\Lambda$ baryon (${\bar \Lambda} = m_{\Lambda_Q}-m_Q$)
and because of the dependence on the heavy quark masses,
$\hat{\xi}_{cb}$ is no longer a universal form factor. The function
$\chi^{\rm ren}(w)$ arises from  higher--dimension operators in the
HQET Lagrangian, and vanishes at zero recoil. Both functions
$\hat{\xi}_{cb}$ and $\xi^{\rm ren}$ are normalized to one at
zero recoil. The numerical values of the correction factors $N_i,
N_i^5$ depend on the value of $\bar{\Lambda}$, which is not precisely
known. We reproduce here (Table~\ref{tab:nw})  Table 4.1 of
Ref.~\cite{HQET2}, where these correction factors are given for all
baryon velocity transfer $w$ accessible in the $\Lambda_b \to \Lambda_c
l{\bar \nu}_l$ decay\footnote{Note the values for those correction factors are somewhat different from the
ones quoted in Ref.~\cite{Neubert94}}.  The parameters ${\bar \Lambda}/2m_b$  and
${\bar \Lambda}/2m_c$ were set to 0.07 and 0.24, respectively. At zero recoil,
Luke's theorem~\cite{Lu90} protects the quantities $F(w)=\sum_i F_i(w)$ and
$G_1(w)$ from ${\cal O}(1/m_Q)$ corrections
\begin{equation}
F(1)=\sum_i F_i(1) = \eta_V+{\cal O}(1/m_Q^2),  \quad
G_1(1)= \eta_A+{\cal O}(1/m_Q^2) 
\end{equation}
where $\eta_V$ and $\eta_A$ are entirely determined by short distance
corrections (ie, $N_1^5(1) = \hat{C}^5_1 (1)$ and $ \sum_i N_i(1) =
\sum_i \hat{C}_i(1)$) which are in principle well known, since they are computed
using perturbative QCD techniques. The second relation might be used
to extract a model independent (up to $1/m_Q^2$ corrections) value of
$|V_{cb}|$ from the measurement of semileptonic $\Lambda_b$ decays
near zero recoil, where the rate is governed by the form factor
$G_1$. From Eq.~(\ref{eq:dg}), one finds
\begin{equation}
\label{limit}
\lim_{w\to 1} \frac{1}{\sqrt{w^2-1}} \frac{{\rm d}\Gamma}{{\rm d}w} =
\frac{G^2 |V_{cb}|^2}{4\pi^3}m_{\Lambda_c}^3 (m_{\Lambda_b} -
  m_{\Lambda_c})^2 \eta_A^2 + {\cal O}(1/m_c^2) 
\end{equation}
\begin{table}[t]
\begin{center}
\begin{tabular}{c|cccc|ccc}\hline\tstrut
$w$  & $N_1$ & $N_2$   & $N_3$ & $\sum_i N_i$  & $N_1^5$ & $N_2^5$ & $N_3^5$\\\hline\tstrut
1.00 & 1.49  & $-$0.36 & $-$0.10 &1.03 & 0.99    & $-$0.42 & 0.15 \\
1.11 & 1.40  & $-$0.32 & $-$0.09 &0.99 & 0.94    & $-$0.37 & 0.13 \\
1.22 & 1.32  & $-$0.30 & $-$0.09 &0.93 & 0.91    & $-$0.34 & 0.12 \\
1.33 & 1.26  & $-$0.27 & $-$0.08 &0.91 & 0.88    & $-$0.31 & 0.11 \\
1.44 & 1.20  & $-$0.25 & $-$0.07 &0.88 & 0.85    & $-$0.28 & 0.10 \\\hline
\end{tabular}
\end{center}
\caption{\footnotesize Correction factors (taken from Ref.~\protect\cite{HQET2}) for
  the $\Lambda_b \to \Lambda_c$ decay form factors}
\label{tab:nw}
\end{table}

\section{Results}
\label{sec:res}
To obtain the wave functions for the $\Lambda_Q$ and $\Xi_Q$ baryons, 
we will use different NRCQM interactions whose details can be found 
in Refs.~\cite{Al04}. Following the
notation of this reference, we will refer to them as AL1, AL1$\chi$,
AL2, AP1, AP2 and BD. Their free parameters had been adjusted in
the meson sector~\cite{BD81,Si96,BFV99}. The potentials considered
differ in the form factors used for the hyperfine terms, the power of
the confining term\footnote{The force which
confines the quarks is still not well understood, although it is
assumed to come from long-range non-perturbative features of
QCD~\cite{Su95}.}  ($p=1$, as suggested by lattice QCD
calculations~\cite{GM84}, or $p=2/3$ which for mesons gives the
correct asymptotic Regge trajectories~\cite{Fabre88}), or the use of a
form factor in the One Gluon Exchange (OGE) Coulomb
potential~\cite{Ru75}. All of them provide
reasonable and similar masses and static properties for $\Lambda_Q,
\Sigma_Q, \Sigma^*_Q, \Xi^\prime, \Xi^\prime_Q, \Xi^*_Q, \Omega_Q$ and
$\Omega^*_Q$ baryons~\cite{Al04}.  

For the $\Lambda_b-$decay we will pay an special attention to the AL1
and AL1$\chi$ inter--quark potentials. The AL1 potential is based on a
phenomenological inter--quark interaction which includes a term with a
shape and a color structure determined from the OGE contribution, and
a confinement potential.  The second model (AL1$\chi$) includes the
same heavy quark--light quark potential as the AL1 model, while the
light quark--light quark is built from the SU(2) chirally inspired
quark-quark interaction of Ref.~\cite{Fe93} which includes a pattern
of spontaneous chiral symmetry breaking, and that was applied with
great success to the meson sector in Ref.~\cite{BFV99},

From the experimental side, the
$\Lambda_b$ semileptonic branching fraction into the exclusive
semileptonic mode was measured in DELPHI to be~\cite{HB-exp}
\begin{equation}
Br(\Lambda_b^0 \to \Lambda_c^+ l^- {\bar \nu}_l) = \left (
5.0^{+1.1}_{-0.8}\, ({\rm stat})\, ^{+1.6}_{-1.2} ({\rm syst}) \right)
\label{eq:br}
\end{equation}
A remark is in order here, the perturbative QCD corrections have been
neglected in Ref.~\cite{HB-exp}, i.e. the correction
factors $N_i, N_i^5$ are computed with ${\hat C}_1={\hat C}_1^5=1$ and 
${\hat C}_{2,3}={\hat C}_{2,3}^5=0$, and a functional form of the 
type 
\begin{equation}
\hat{\xi}_{cb}(w)= e^{-{\hat \rho}^2(w-1)}\label{eq:exp}
\end{equation}
is also assumed in that reference, where it is also found that\footnote{Note
that $-{\hat \rho}^2$ is not the slope at the origin of the universal Isgur-Wise
function $\xi^{ren}(w)$ introduced in Eq.(\ref{eq:defxi}).}
\begin{equation}
{\hat \rho}^2 = 2.0^{+0.8}_{-1.1} \label{eq:rho_exp}
\end{equation}
where all uncertainties quoted in Ref.~\cite{HB-exp} have been added
in quadratures. On the other hand, the branching fraction given by the Particle
Data Group is~\cite{pdg04}  
\begin{equation}
\label{eq:brpdg}
Br(\Lambda_b^0 \to \Lambda_c^+ l^- {\bar \nu}_l+{~\rm anything}) = \left (
9.2 \pm 2.1 \right)\% 
\end{equation}
which is hardly consistent to that quoted in
Eq.~(\ref{eq:br}). Nevertheless, none of the values quoted in
Eqs.~(\ref{eq:br}) and~(\ref{eq:brpdg}) correspond to
direct measurements. We will assume here, an error weighted averaged
value\footnote{We add in quadratures the statistical and systematic
uncertainties quoted in Eq.~(\ref{eq:br}).}  of those given in
Eqs.~(\ref{eq:br}) and~(\ref{eq:brpdg})
\begin{equation}
\langle Br(\Lambda_b^0 \to \Lambda_c^+ l^- {\bar \nu}_l) \rangle_{\rm
  avg} = \left (
6.8 \pm 1.3 \right)\% 
\end{equation}\label{eq:bravg}
The total $\Lambda_b^0$ width is given by its lifetime
$\tau_{\Lambda_b^0} = 1.229 \pm 0.080$ ps~\cite{pdg04} and thus one finds
\begin{equation}
\Gamma(\Lambda_b^0 \to \Lambda_c^+ l^- {\bar \nu}_l) = \left (
5.5 \pm 1.4 \right) 10^{10} s^{-1} \label{eq:sml}
\end{equation}
Besides, data from $Z$ decays in DELPHI have been searched for ${\bar
  {\rm B}^0_{\rm d}} \to {\rm D}^{*+}l^-{\bar \nu}_l$ decays. These events are
  used to measure the CKM matrix element $|V_{cb}|$~\cite{Exp3}
\begin{equation}
|V_{cb}| = 0.0414 \pm 0.0012\,({\rm stat})\,\pm 0.0021\,({\rm syst})\,
 \pm 0.0018\,({\rm theory})\, \label{eq:vcb}
\end{equation}

Let us first examine  the bare NRCQM predictions without including HQET
constraints.

\subsection{NRCQM Form Factors} 
\label{sec:nrcqmff}

\begin{figure}
\centerline{\includegraphics[height=20cm]{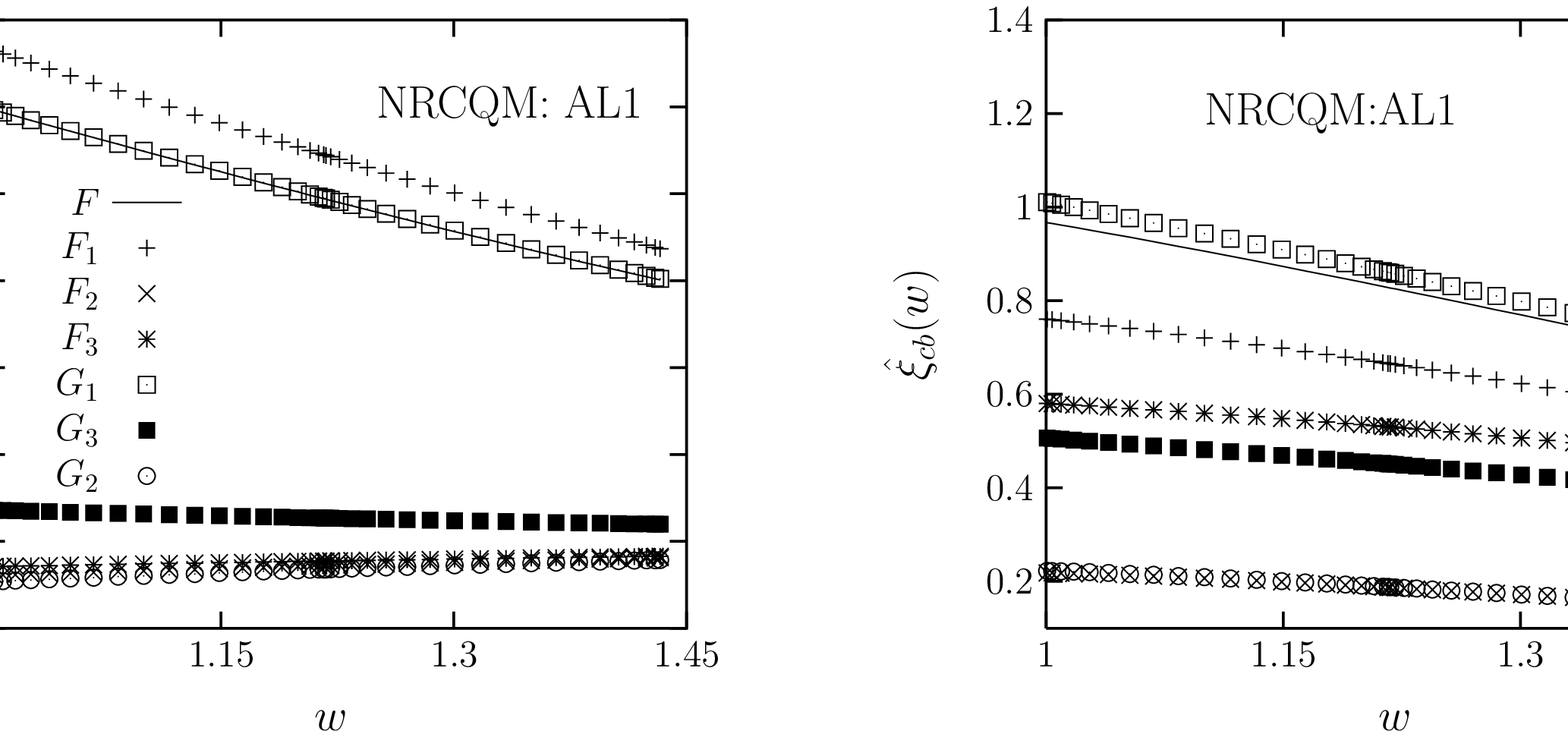}}
\vspace{-12cm}
\caption{\footnotesize  NRCQM $\Lambda_b \to \Lambda_c$ form factors (left)
  and $\hat{\xi}_{cb}$ function (right) from the AL1 inter--quark interaction.}
  \label{fig:nrcqm}
\end{figure}

In Fig.~\ref{fig:nrcqm} we present the $\Lambda_b \to \Lambda_c$ form
factors obtained from the AL1 inter--quark interaction (left) and also
the predictions for the $\hat{\xi}_{cb}$ function (right) as extracted
from any of the form factors ($\hat{\xi}_{cb} = F_i/N_i, G_i/N_i^5,
F/\sum_iN_i, \, i=1,2,3$) shown in the left panel. The correction
factors $N_i, N_i^5$ are taken from Table~\ref{tab:nw}. Several
comments are in order:
\begin{itemize}
\item As expected from HQS, the form factors $F_2$, $F_3$, $G_2$ and $G_3$ 
  are significantly smaller than the dominant ones $F_1$ and
  $G_1$.
\item Recalling the discussion of Subsect.~\ref{sec:me} on vector
  current conservation for degenerate transitions, one must conclude
  that the NRCQM predictions for the $F_2$ and $F_3$ form factors are
  not reliable at all, since their sizes are comparable to the expected  
  theoretical uncertainties, ${\cal O}\left(1-m_{\Lambda_Q}/M_{\rm
  tot}\right)$, affecting them. Presumably, one should draw similar 
  conclusions for the axial $G_2$ and $G_3$ form factors. As clearly seen in 
  Fig.~\ref{fig:nrcqm} the $\hat{\xi}_{cb}(w)$ functions obtained from the 
  $F_2$, $F_3$, $G_2$ and $G_3$
  form factors  substantially differ among themselves and are in
  complete disagreement to those obtained from the $F$ and $G_1$ form
  factors.
  
\item NRCQM predictions for the vector $F$ and axial $G_1$ form
  factors are much more reliable, and lead to similar $\hat{\xi}_{cb}$
  functions, with discrepancies smaller than around 4\%. Such
  discrepancies can be attributed either to ${\cal O}(1/m_Q^2)$
  corrections, not included in $\hat{\xi}_{cb}$, or to deficiencies of
  the NRCQM. Lattice results of Ref.~\cite{HB-Lattice} for these two
  form factors, though have large errors, are in good agreement with
  the results shown in Fig.~\ref{fig:nrcqm}.
 
\end{itemize}

\subsection{HQET and NRCQM Combined Analysis.}

To improve the NRCQM results, we proceed as follows. We assume the NRCQM 
estimate
of the vector form factor $F$ ($F=F_1+F_2+F_3$) to be
correct for the whole range of
velocity transfers accessible in the physical decay\footnote{ Let us 
remind here, that the
NRCQM gives correctly $F(1)$ in the case of degenerate transitions.}, and use it to
obtain the flavor depending $\hat{\xi}_{cb}$ function. Now by using
Eq.~(\ref{eq:deff}) and the HQET coefficients $N_i, N_i^5$ compiled in
Table~\ref{tab:nw}, we reconstruct the rest of form factors, in terms
of which we can predict the longitudinal and transverse differential
decay widths and the asymmetry parameters defined in
Subsect.~\ref{sec:cont}. We will estimate the theoretical error of the
present analysis by accounting for the spread of the results obtained
when all calculations are repeated by determining $\hat{\xi}_{cb}$
from the NRCQM $G_1$ form factor and/or by using different
inter--quark interactions.

\subsubsection{$\Lambda_b$ Decay}

\begin{figure}
\centerline{\includegraphics[height=20cm]{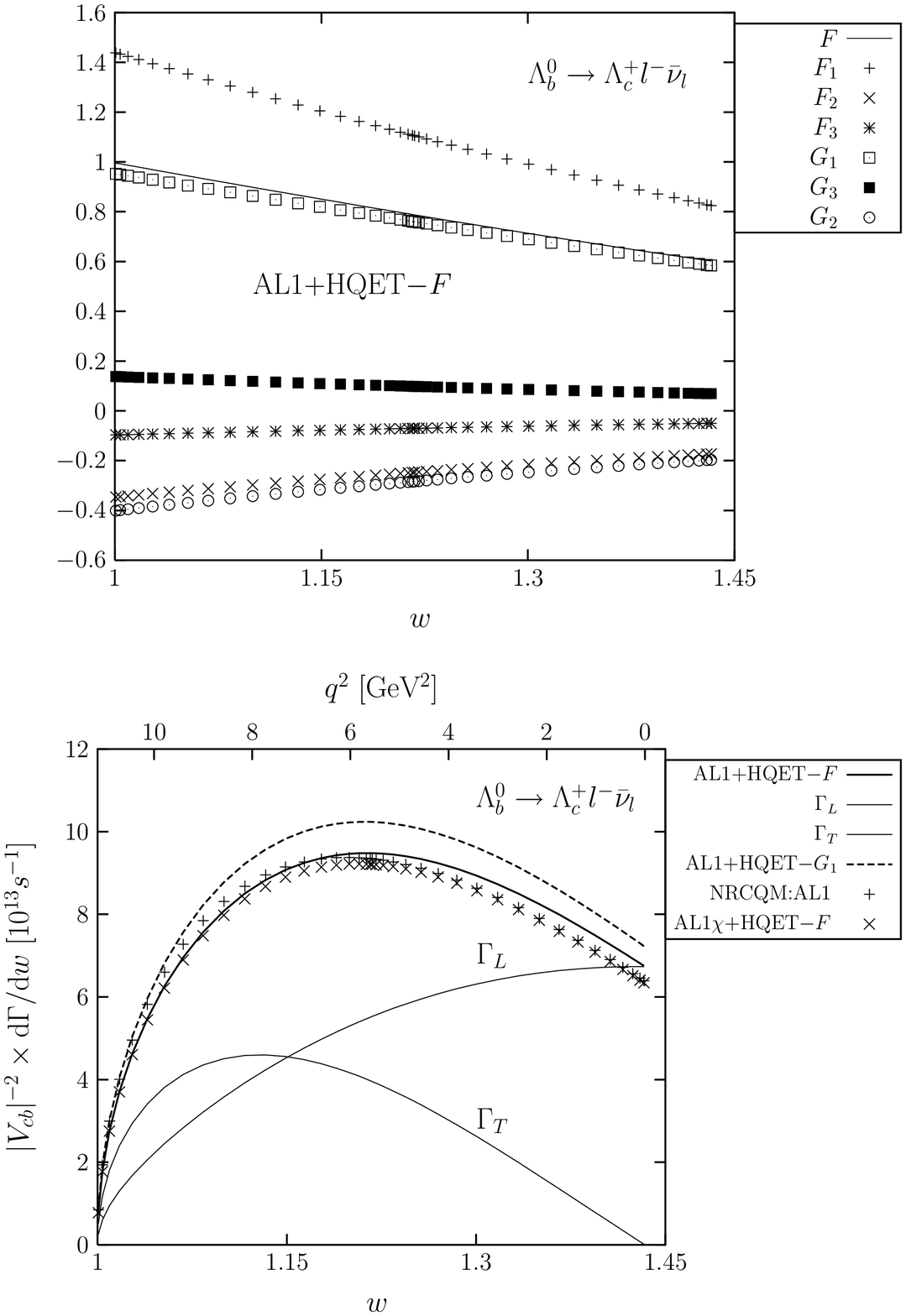}}
\caption{ \footnotesize $\Lambda_b \to \Lambda_c$ form factors (top)
  and differential decay width (bottom) constrained by means of the
  HQET relations of Eq.~(\ref{eq:deff}). AL1 and AL1$\chi$
  inter--quark interactions have been used, and HQET$-F$ and
  HQET$-G_1$ stand for HQET models where $\hat{\xi}_{cb}$ is
  determined from the NRCQM predictions for the $F$ and $G_1$ form
  factors, respectively. For the AL1+HQET$-F$ model, longitudinal and
  transverse differential decay distributions are plotted as well. For
  comparison in the bottom panel, we also show the bare NRCQM decay
  width distribution obtained from the form factors plotted in the
  left panel of Fig.~\protect\ref{fig:nrcqm}. }
  \label{fig:lambda}
\end{figure}
Results of our HQET improved NRCQM analysis for the $\Lambda_b$ decay
are compiled in Fig.~\ref{fig:lambda} and Tables~\ref{tab:lambda}
and~\ref{tab:asy}. In the first of the tables, we give the total and
partially integrated semileptonic decay widths, split  into the
contributions to the rate from transversely $(\Gamma_T)$ and
longitudinally $(\Gamma_L)$ polarized $W$'s, and the value of the
flavor depending $\hat{\xi}_{cb}(w)$ function and its derivatives at
zero recoil, together with our estimates for the uncertainties of the
present analysis. We also compare, when possible, with the lattice
results of Ref.~\cite{HB-Lattice}. In the second of the tables, we
compile our predictions for the $w-$averaged asymmetry parameters
defined in Eqs.~(\protect\ref{eq:at})--(\protect\ref{eq:rlt}). Our
results compare exceptionally well to those obtained by Cardarelli and
Simula from a light--front constituent quark model~\cite{HB-LFQM}. On
the other hand, we should mention that the NRCQM described in
Subsect.~\ref{sec:nrcqmff} leads to similar (discrepancies of around
2-3\%) differential decay rates, as can be appreciated in
Fig.~\ref{fig:lambda}. From the discussion in
Subsect.~\ref{sec:nrcqmff}, this fact should be considered as an
accident. For the $w-$averaged asymmetry parameters given in
Table~\ref{tab:asy}, discrepancies are in general higher, being of the order of
20\% for the $\langle a_T \rangle$ and $\langle \alpha^\prime
\rangle$ asymmetries.
\begin{table}[ht]
\begin{center}
\begin{tabular}{r|ccccccc|cc}
                   & HQET$-F$~ & HQET$-G_1$ ~& HQET$-F$~  &
                   HQET$-F$~ &  HQET$-F$~ & HQET$-F$~  & HQET$-F$~  & Theor.
                   &Lattice\\
                   & AL1      &  AL1       & AL1$\chi$ & AL2 &
                   AP1 & AP2& BD & Avg.     & Ref.~\protect\cite{HB-Lattice}
                   \\\hline\tstrut 
 $\Gamma$~~ &  3.46    &    3.73    &    3.35   & 3.57 & 3.50 &
                   3.60 & 3.49 & $3.46^{+0.27}_{-0.11}$&$-$\\\jtstrut 
$\Gamma_L$ & 2.14   &2.31        & 2.07      &2.22  &2.18  & 2.25
                      & 2.16 & $2.14^{+0.17}_{-0.07}$            &$-$\\\jtstrut
$\Gamma_T$ &1.31      & 1.42       & 1.28      &1.34  &1.33  & 1.36
                      & 1.32 & $1.31^{+0.11}_{-0.03}$            &$-$\\\hline\tstrut 
${\hat \Gamma}_L$, $w_0$~   & & & & &&&&&\\\tstrut  
1.10~ &0.23 &0.25 &0.22 &0.22&0.23&0.23&0.23&$0.23^{+0.02}_{-0.01}$
                   &$0.23^{+0.03}_{-0.02}$\\\jtstrut  
1.15~ &0.43 & 0.47&0.42 &0.43&0.43&0.43&0.43&$0.43^{+0.04}_{-0.01}$ &$0.44^{+0.08}_{-0.06}$\\\jtstrut  
1.20~ & 0.68& 0.73&0.66 &0.69&0.68&0.68&0.68&$0.68^{+0.05}_{-0.02}$ &$0.71^{+0.17}_{-0.13}$\\\jtstrut  
1.25~ & 0.96&1.04 &0.94 &0.98&0.97&0.97&0.96&$0.96^{+0.08}_{-0.02}$
                   &$1.0^{+0.3}_{-0.2}$\\\jtstrut  
1.30~ & 1.26&1.37 &1.23 &1.29&1.28&1.28&1.27&$1.26^{+0.11}_{-0.03}$ &$1.4^{+0.5}_{-0.4}$\\\jtstrut  
1.35~ &1.59 &1.71 &1.54 &1.63&1.61&1.61&1.60&$1.59^{+0.12}_{-0.05}$
                   &$-$\\\hline
${\hat \Gamma}_T$, $w_0$~   & & & & &&&&&\\\tstrut  
1.10~ &0.34 &0.37 &0.34 &0.34&0.34&0.34&0.34&$0.34^{+0.03}_{-0.00}$
                   &$0.34^{+0.06}_{-0.04}$\\\jtstrut  
1.15~ &0.57 & 0.62&0.56 &0.58&0.57&0.57&0.57&$0.57^{+0.05}_{-0.01}$ &$0.53^{+0.16}_{-0.14}$\\\jtstrut  
1.20~ & 0.79& 0.86&0.78 &0.80&0.80&0.80&0.79&$0.79^{+0.07}_{-0.01}$ &$0.7^{+0.3}_{-0.3}$\\\jtstrut  
1.25~ & 0.98&1.06 &0.96 &1.00&0.99&0.99&0.99&$0.98^{+0.08}_{-0.02}$ &$0.8^{+0.6}_{-0.5}$\\\jtstrut  
1.30~ & 1.14&1.23 &1.11 &1.16&1.15&1.15&1.14&$1.14^{+0.09}_{-0.03}$ &$0.8^{+0.9}_{-0.8}$ \\\jtstrut  
1.35~ &1.24 &1.35 &1.22 &1.27&1.26&1.26&1.25&$1.24^{+0.11}_{-0.02}$ &$-$\\\hline \jtstrut 
$\hat{\xi}_{cb}(1)$ &0.97 &1.01 &0.97&0.97 & 0.97 & 0.97 & 0.97
                   &$0.97^{+0.04}_{-0.00}$ & $0.99\pm 0.01$\\\jtstrut 
$-\hat{\xi}_{cb}^\prime(1)$ & $0.58$&$0.65$ &$0.64$&0.52&0.58 &
                   0.52 & 0.56 &$0.58^{+0.07}_{-0.06}$ & $1.1\pm 1.0$\\\jtstrut 
$-\hat{\xi}_{cb}^{\prime\prime}(1)$ & $0.73$&$0.59$ &$0.72$ &0.79&0.63&0.70&0.82&$0.73^{+0.09}_{-0.14}$ &$-$\\\jtstrut 
$\hat{\xi}_{cb}^{\prime\prime\prime}(1)$ &2.3 & 2.0&2.6 & 2.3&1.8&1.9&2.5&$2.3^{+0.3}_{-0.5}$&$-$\\\hline
\end{tabular}
\end{center}
\caption{ \footnotesize $\Lambda_b$ semileptonic decay: Theoretical
  predictions for totally and partially (${\hat
  \Gamma}_{L,T}=\int_1^{w_0} {\rm d}w \frac{{\rm d} \Gamma_{L,T}}{{\rm
  d}w}$) integrated decay widths, in units of $|V_{cb}|^2 10^{13}$
  s$^{-1}$, and for $\hat{\xi}_{cb}(w)$ and its derivatives at zero
  recoil. The meaning of columns 2 to 8 is the same as in
  Fig.~\protect\ref{fig:lambda}, with the obvious changes due to the
  use of different inter--quark interactions. In the ninth column
  (Theor. Avg.) we give our final results with theoretical
  uncertainties obtained from the spread of the results shown in the
  table. Finally in the last column we compile the Lattice QCD results
  of Ref.~\protect\cite{HB-Lattice}. }\label{tab:lambda}
\end{table}
\begin{table}[htb]
\begin{center}
\begin{tabular}{c|ccccccc}
& $R_{L/T}$ & $\langle a_T \rangle$  & $\langle a_L \rangle$ &
  $\langle P_L \rangle$ & $\langle \alpha^\prime \rangle$    &
  $\langle \alpha^{\prime\prime} \rangle$ & $\langle \gamma \rangle$\\\hline\tstrut        
$\Lambda_b^0 \to \Lambda_c^+ l^- {\bar \nu}_l$~ & $1.63\pm 0.02$~ &
  $-0.665\pm 0.002$~ &  $-0.954\pm 0.001$~ & $-0.844\pm 0.003$~ &  $-0.156\pm
  0.001$~ & $-0.531\pm 0.004 $~ & $0.439\pm 0.004$ \\\jtstrut
$\Xi_b^0 \to
\Xi_c^+ l^- {\bar \nu}_l$~ & $1.53\pm 0.04$~ &
  $-0.628\pm 0.004$~ &  $-0.945\pm 0.002$~ & $-0.820\pm 0.004$~ &  $-0.154\pm
  0.001$~ & $-0.508\pm 0.008 $~ & $0.475\pm 0.006$ \\\hline       
\end{tabular}
\end{center}
\caption{ \footnotesize Theoretical predictions for the $w-$averaged
  asymmetry parameters defined in
  Eqs.~(\protect\ref{eq:at})--(\protect\ref{eq:rlt}). We quote central
  values from the AL1+HQET$-F$ model and the theoretical
  uncertainties have been determined as in Table~\protect\ref{tab:lambda}.
  }\label{tab:asy}
\end{table}

\noindent
From our theoretical determination of the total semileptonic width in
Table~\ref{tab:lambda} and the experimental estimate in
Eq.~(\ref{eq:sml}), we get
\begin{equation}
|V_{cb}| = 0.040\pm 0.005~({\rm stat})~^{+0.001}_{-0.002}~({\rm
 theory})
\label{eq:ourvcb}
\end{equation} 
in remarkable agreement with the recent determination of this
parameter from ${\bar {\rm B}^0_{\rm d}} \to {\rm D}^{*+}l^-{\bar
\nu}_l$ decays (Eq.~(\ref{eq:vcb})). The experimental uncertainties on
the $\Lambda_b$ semileptonic branching ratio turn out to be the major
source of error in the present determination of $|V_{cb}|$, being the
theoretical error in both Eqs.~(\ref{eq:vcb}) and~(\ref{eq:ourvcb})
comparable in size.  We point out nevertheless that our determination
of $|V_{cb}|$ is based on a NRCQM description of the baryon and as
such it is not model independent.  From a conceptual point of view a
determination of $|V_{cb}|$ based on Eq.~(\ref{limit}) would be
preferred in a non-relativistic approach as both baryons are at rest
in the $w\to 1$ limit. Unfortunately the lack of enough experimental
data in that region prevents such a calculation.  Note nevertheless
that our partially integrated width ($\hat{\Gamma}_L+\hat{\Gamma}_T$)
is in good agreement with the lattice results of
Ref.~\cite{HB-Lattice} for $w$ values up to $w\approx 1.20$
($|\vec{q}\,|\approx 0.66 m_{\Lambda_c}$) where lattice calculations
are reliable, and that our total width
$\Gamma=3.46^{+0.27}_{-0.11}\,|V_{cb}|^2\,10^{13}$s$^{-1}$ agrees with
the value of $\Gamma=3.1\pm 1.0\,|V_{cb}|^2\,10^{13}$s$^{-1}$ obtained
in Ref.~\cite{HB-SR3} using QCD sum rules.
  
\noindent 
With respect to the
$1/m_Q-$corrected Isgur-Wise function $\hat{\xi}_{cb}(w)$  our results show
 a clear
departure\footnote{Note for instance that
$\hat{\xi}_{cb}^{\prime\prime}(1)$ and
$\hat{\xi}_{cb}^{\prime\prime\prime}(1)$ have changed signs with
respect those deduced from Eq.(\ref{eq:exp}).} from a single
exponential (see Eq.(\ref{eq:exp})) functional form, and
instead, in the velocity transfer range accessible in the physical
decay, it is rather well described by a rank three polynomial in
powers of $(w-1)$. In what $\hat{\xi}_{cb}^{\prime}(1)$ respects, our
estimate lies in the lower end of the range of
Eq.~(\ref{eq:rho_exp}). As mentioned above, the perturbative QCD
corrections were neglected in Ref.~\cite{HB-exp}. If we do not include
the short distance contributions when relating the NRCQM AL1 $F$ form
factor and the HQET $\hat{\xi}_{cb}(w)$ function, the slope of this
latter function becomes larger (in absolute value), ie
$\hat{\xi}_{cb}^{\prime}(1)=-0.99$, in closer agreement with the DELPHI
estimate. Besides, the assumption in Ref.~\cite{HB-exp} of the
functional form of Eq.~(\ref{eq:exp}) leads also to larger, in
absolute value, slopes. Thus, to get a semileptonic decay width of 3.46
$|V_{cb}|^2 10^{13}$s$^{-1}$  (our
prediction for AL1+HQET-$F$), a value of ${\hat \rho}^2 = 1.20$ is required\footnote{Note
also that both approaches provide ${\rm d}\Gamma/{\rm d}w$
distributions which are quite similar making it difficult for experimentalists 
 to decide which one is preferred.}.

Finally, we would like to remark the minor differences, of the order
of a few per cent, existing between the AL1 and AL1$\chi$ NRCQM based
predictions for the decay distributions. This is a common feature,  when
the different inter--quark interactions studied in Ref.~\cite{Al04} are
considered. All results are  compiled in Table~\ref{tab:lambda}.

\subsubsection{$\Xi_b$ Decay}

\begin{figure}
\centerline{\includegraphics[height=20cm]{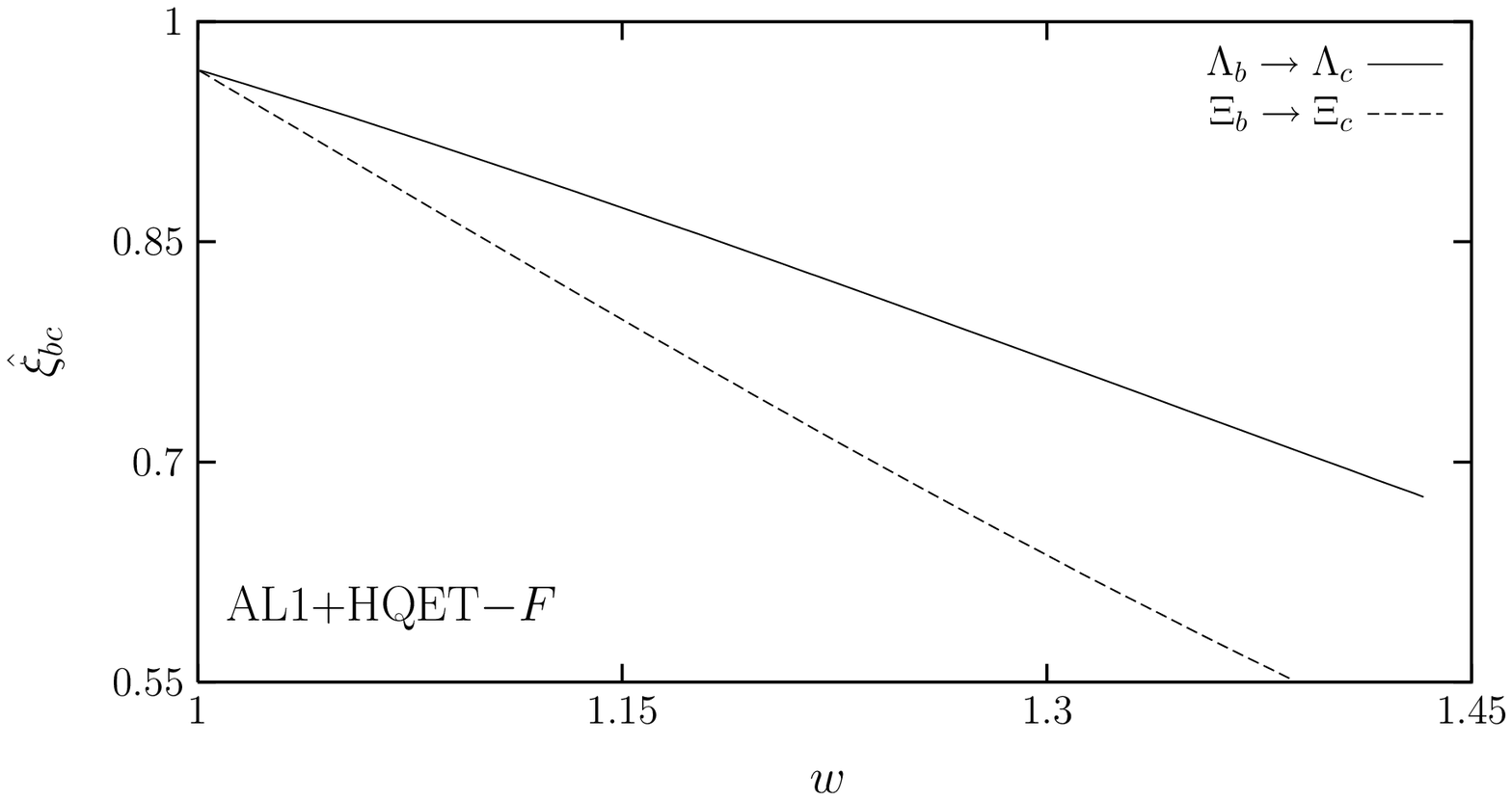}}
\vspace{-12.cm}
\caption{ \footnotesize $\Lambda_b \to \Lambda_c$ and $\Xi_b \to
  \Xi_c$ $1/m_Q-$corrected Isgur-Wise functions, $\hat{\xi}_{cb}(w)$,
  from the AL1+HQET$-F$ model.  }
  \label{fig:iw}
\end{figure}

\begin{table}[ht]
\begin{center}
\begin{tabular}{r|cccccc|cc}
                   & HQET$-F$~ & HQET$-G_1$ ~& HQET$-F$~  &
                   HQET$-F$~ & HQET$-F$~ & HQET$-F$~ 
                   & Theor. &Lattice\\
                   & AL1      &  AL1       & AL2 & AP1 & AP2 &
                   BD & Avg.    & Ref.~\protect\cite{HB-Lattice} \\\hline\tstrut 
 $\Gamma$~~&  2.96    &    3.21    & 3.04  &  2.97&  3.12& 3.08
                   &$2.96^{+0.25}_{-0.00}$ & $-$\\\jtstrut 
 $\Gamma_L$~~&  1.79    &    1.94    & 1.85  &  1.80&  1.90& 1.88
                   &$1.79^{+0.15}_{-0.00}$ &$-$\\\jtstrut 
 $\Gamma_T$~~&  1.17    &    1.27    & 1.19  &  1.17&  1.22& 1.21
                   &$1.17^{+0.10}_{-0.00}$ &$-$\\\hline
${\hat \Gamma}_L$, $w_0$~   & & & & & & &  \\\tstrut  
1.10~ &0.27 &0.29 & 0.27&0.27 &0.27 &0.27&$0.27^{+0.02}_{-0.00}$ &$0.28^{+0.02}_{-0.03}$\\\jtstrut  
1.15~ &0.49 & 0.54&0.50 &0.49 &0.50 &0.50  &$0.49^{+0.05}_{-0.00}$ &$0.54^{+0.07}_{-0.08}$
                   \\\jtstrut  
1.20~ & 0.75& 0.82& 0.76&0.75 &0.77 &0.77 &$0.75^{+0.07}_{-0.00}$ &$0.86^{+0.14}_{-0.16}$\\\jtstrut  
1.25~ & 1.03&1.12 &1.05 & 1.03&1.07 &1.06 &$1.03^{+0.09}_{-0.00}$ &$1.2^{+0.2}_{-0.3}$\\\jtstrut  
1.30~ &1.31 &1.42 &1.34 & 1.31& 1.37&1.36 &$1.31^{+0.11}_{-0.00}$ &$1.7^{+0.4}_{-0.4}$\\\hline 
${\hat \Gamma}_T$, $w_0$~   & & & & & & &  \\\tstrut  
1.10~ &0.39 &0.43 & 0.39&0.39 &0.40 &0.40&$0.39^{+0.04}_{-0.00}$ &$0.38^{+0.05}_{-0.05}$\\\jtstrut  
1.15~ &0.63 & 0.68&0.63 &0.63 &0.64 &0.64  &$0.63^{+0.05}_{-0.00}$ &$0.58^{+0.13}_{-0.15}$
                   \\\jtstrut  
1.20~ & 0.83& 0.90& 0.84&0.83 &0.85 &0.85 &$0.83^{+0.07}_{-0.00}$ &$0.7^{+0.3}_{-0.3}$\\\jtstrut  
1.25~ & 0.99&1.08 &1.01 & 0.99&1.02 &1.02 &$0.99^{+0.09}_{-0.00}$ &$-$\\\jtstrut  
1.30~ &1.10 &1.19 &1.12 & 1.10& 1.14&1.13 &$1.10^{+0.09}_{-0.00}$ &$-$\\\hline \jtstrut 
$\hat{\xi}_{cb}(1)$ &0.97 &1.01 &0.96     & 0.97& 0.97& 0.97
                    &$0.97^{+0.04}_{-0.01}$ & $0.99\pm 0.01$\\\jtstrut 
$-\hat{\xi}_{cb}^\prime(1)$ & $1.14$&$1.22$ & 1.06     & 1.15&
                   1.02& 1.05                     & $1.14 \pm 0.08$ &$1.4\pm 0.8$\\\jtstrut 
$-\hat{\xi}_{cb}^{\prime\prime}(1)$ & $0.13$&$0.00$ &0.30 &
                   $0.03$&0.30 &0.37&$0.13^{+0.24}_{-0.16}$ &$-$\\\jtstrut 
$\hat{\xi}_{cb}^{\prime\prime\prime}(1)$ &4.6 & 4.5&4.7 &4.3  &4.3  &
                   4.7 &$4.6^{+0.1}_{-0.3}$ & $-$\\\hline
\end{tabular}
\end{center}
\caption{ \footnotesize As in Table~\ref{tab:lambda}, but for $\Xi_b$
  baryon semileptonic decay.}\label{tab:xi}
\end{table}
Results of our HQET improved NRCQM analysis for the $\Xi_b$ decay are
compiled in Tables~\ref{tab:asy} and~\ref{tab:xi}. As in the
$\Lambda_b-$decay case, the decay parameters do not depend
significantly on the potential, among those considered in this
work. This fact allows us to make precise theoretical predictions,
which nicely agree to the lattice results of
Ref.~\cite{HB-Lattice}. On the other hand, we find small SU(3)
deviations, and thus as a matter of example we find
\begin{equation}
\frac{\Gamma\left (\Xi_b \to \Xi_c\, l {\bar
    \nu}_l\right)}{\Gamma\left (\Lambda_b \to \Lambda_c\, l 
{\bar \nu}_l\right)} = 0.86^{+0.10}_{-0.07},
\end{equation}
which will naturally fit within  SU(3) symmetry expectations. 

Finally, in Fig.~\ref{fig:iw} we plot the $\Lambda_b \to \Lambda_c$
and $\Xi_b \to \Xi_c$ $1/m_Q-$corrected Isgur-Wise functions,
$\hat{\xi}_{cb}(w)$, from the AL1+HQET$-F$ model. We see there, the
size of possible SU(3) symmetry violations as a function of the
velocity transfer $w$. The zero recoil slope, in absolute value, is
significantly larger for the $\Xi_b \to \Xi_c$ transition than for the
the $\Lambda_b \to \Lambda_c$ one.  A similar behavior is also found
in the meson sector in the B$\to {\rm D}_{(s)},{\rm D}^*_{(s)}$
decays.  Lattice
calculations show that the slope at zero recoil of the mesonic
Isgur-Wise function is larger in magnitude in the case where the
spectator quark is a strange one~\cite{UKQCD1}.

\section{Concluding Remarks} 
\label{sec:concl}

We have identified two of the main deficiencies of the NRCQM
description of the semileptonic decay of the $\Lambda_b$ and $\Xi_b$
baryons: i) A standard momentum expansion of the electroweak current
is totally unappropriated, far from the zero recoil point. ii) Within
the usual spectator model approximation, with only one--body current
operators, the vector part of the electroweak charged current is not
conserved for degenerate transitions. Both drawbacks prevent NRCQM's
to make reliable predictions of form factors and totally integrated
decay rates. In the present work we have solved both deficiencies, and
thus we have developed a novel expansion for the electroweak current
operator, where all orders on the transferred momentum $\vec{q}$ are
kept. To improve on the second of the mentioned deficiencies, we have
also implemented HQET constraints among the form-factors. In addition
to other desirable features, we would restore in this way, vector
current conservation for degenerate transitions.

Our HQET improved NRCQM analysis leads to an accurate and reliable
description of the $\Lambda_b$ semileptonic decay. Thus,
we determine the $1/m_Q-$corrected Isgur-Wise function which governs
this process and, thanks to the branching fraction values quoted in 
Refs.~\cite{pdg04} and~\cite{HB-exp}, extract the modulus of 
the $cb$ CKM matrix element
(Eq.~(\ref{eq:ourvcb})). Our determination of $|V_{cb}|$ comes out in
total agreement with that obtained from semileptonic ${\rm B}\to {\rm
D}^*$ decays (Eq.~(\ref{eq:vcb})), and if it suffers from larger
uncertainties that the latter one is because of a poorer experimental
measurement of the semileptonic branching fraction for the $\Lambda_b$
case. We also give various $w-$averaged asymmetry parameters, which
determine the angular distribution of the decay.

In what respects to the $\Xi_b-$semileptonic decay, we also find an
accurate and reliable description of the various physical magnitudes
which govern this transition, and  find SU(3) symmetry deviations
 of the order of 15\%. At zero recoil, the  $1/m_Q-$corrected
 Isgur-Wise function slope, in absolute value, is significantly larger
for the $\Xi_b \to \Xi_c$ transition than for the the $\Lambda_b \to
\Lambda_c$ one.

\appendix
\section{Evaluation of the ${\cal I}$ and ${\cal K}$ Integrals}

We use a partial wave expansion of the $\Lambda_b$, $\Lambda_c$,
$\Xi_b$ and $\Xi_c$  wave functions, 
\begin{equation}
\Psi^{\Lambda_Q}_{ll}(r_1,r_2,r_{12}) = \sum_{l=0}^{+\infty}
f^Q_l(r_1,r_2) P_l(\mu), \quad 
\Psi^{\Xi_Q}_{ls}(r_1,r_2,r_{12}) = \sum_{l=0}^{+\infty}
g^Q_l(r_1,r_2) P_l(\mu), \quad Q=c,b \\
\end{equation}
where $\mu$ is the cosine of the angle between the vectors $\vec
{r}_1$ and $\vec {r}_2$, being $r_{12}=(~r_1^2+r_2^2-2 r_1 r_2
\mu)^\frac12$, and $P_l$ Legendre polynomials of rank $l$. Therefore,
the radial functions $f^Q_l(r_1,r_2)$ and $g^Q_l(r_1,r_2)$ are
obtained from their corresponding wave function by means of:
\begin{equation}
f^Q_l(r_1,r_2) = \frac{2l+1}{2}\int_{-1}^{+1}d\mu P_l(\mu)
\Psi^{\Lambda_Q}_{ll}(r_1,r_2,r_{12}), \quad
g^Q_l(r_1,r_2) = \frac{2l+1}{2}\int_{-1}^{+1}d\mu P_l(\mu)
\Psi^{\Xi_Q}_{ls}(r_1,r_2,r_{12})
\end{equation}
where $r_{12}$ depend on $r_1,r_2$ and $\mu$. In terms of integrals of
the above functions, the baryon factor ${\cal I}^{cb}_{\Lambda}$ reads (we
recall that for $\Lambda_b$ decay, $|\vec{q}\,|=m_{\Lambda_c}
\sqrt{w^2-1}$),
\begin{eqnarray}
{\cal I}^{cb}_{\Lambda} (w)&=& 
(4\pi)^2 \sum_l\sum_{l^\prime}\sum_{l^{\prime\prime}}
(-1)^{l^{\prime\prime}} (ll^\prime l^{\prime\prime}|000)^2 
\int_0^{+\infty} dr_1r_1^2 j_{l^{\prime\prime}}(x_1) \int_0^{+\infty} dr_2r_2^2
j_{l^{\prime\prime}}(x_2)
[f^c_l(r_1,r_2)]^* f^b_{l^\prime}(r_1,r_2)  \label{eq:defI}
\end{eqnarray}
where the flavor of the light quarks ($q,q'$) are {\it up} and {\it down}
\begin{equation}
x_1=\frac{m_q|\vec{q}\,|}{M^c_{\rm tot}}r_1, \quad
x_2= \frac{m_{q^\prime}|\vec{q}\,|}{M^c_{\rm tot}}r_2 \label{eq:defx1x2}
\end{equation}
and $M^c_{\rm tot}=m_u+m_d+m_c$, with $m_u=m_d$. Besides, $(ll^\prime
l^{\prime\prime}|000)$ is a Clebsh-Gordan coefficient and $j_l$ are
spherical Bessel's functions.

On the other hand,  the baryon factor ${\cal K}$  can be computed as
\begin{eqnarray}
{\cal K}^{cb}_\Lambda (w)&=& 
 \frac{16\pi^2}{\sqrt{3}|\vec{q}\,|}
 \sum_l\sum_{l^\prime}\sum_{l^{\prime\prime}}\sum_{l^{\prime\prime\prime}}
 \sum_{L=l^\prime+1, l^\prime-1} (-1)^{l+L}\,{\rm
 i}^{l^{\prime\prime}+l^{\prime\prime\prime}+1 }  
\Big ((2L+1)(2l^{\prime\prime}+1)   
 (2l^{\prime\prime\prime}+1) \Big)^\frac12    (lLl^{\prime\prime}|000)
  \nonumber \\ 
 &\times&
 (l^\prime l l^{\prime\prime\prime}|000) 
(l^{\prime\prime\prime} l^{\prime\prime}1|000) 
W(l^{\prime\prime\prime}l1L;l^\prime l^{\prime\prime})  \int_0^{+\infty}
 dr_1r_1^2 \int_0^{+\infty} dr_2r_2^2 \Big \{ \nonumber  
j_{l^{\prime\prime}}(x_1)  j_{l^{\prime\prime\prime}}(x_2) 
[f^c_l(r_1,r_2)]^*  \\ 
 &\times&  
\Omega_L[f^b_{l^\prime}(r_1,r_2)]
 +j_{l^{\prime\prime}}(\frac{m_{q^\prime}}{m_q}x_1)  
j_{l^{\prime\prime\prime}}(\frac{m_q}{m_{q^\prime}}x_2) 
[f^c_l(r_1,r_2)]^*  \Omega_L[f^b_{l^\prime}(r_1,r_2)] \Big \}
\end{eqnarray}
where $W(...)$ are Racah coefficients, and the differential operators
$\Omega_L$ are defined as
\begin{equation}
\Omega_{L=l^\prime + 1} = - \left ( \frac{l^\prime +1}{2l^\prime
  +1}\right)^\frac12\left[\frac{\partial}{\partial r_1}-
  \frac{l^\prime}{r_1} \right], \quad \Omega_{L=l^\prime - 1} = \left
  ( \frac{l^\prime}{2l^\prime
  +1}\right)^\frac12\left[\frac{\partial}{\partial r_1}+
  \frac{l^\prime+1}{r_1} \right]\label{eq:oemgas}
\end{equation}
Note that ${\cal K}$ remains finite in the limit $|\vec{q}\,|\to 0$,
since one cannot take the orders ($l^{\prime\prime}$ and
$l^{\prime\prime\prime}$) of both Bessel functions to be 0 due to the
Clebsh-Gordan coefficient $(l^{\prime\prime\prime}
l^{\prime\prime}1|000)$. For $\Xi_b$ decay, ${\cal I}^{cb}_{\Xi}$ and
${\cal K}^{cb}_{\Xi}$ are obtained from
Eqs.~(\ref{eq:defI})--(\ref{eq:oemgas}), by replacing $f-$type radial
wave functions by $g-$type ones and taking $m_{q^\prime}=m_s$.

Finally, in the $m_b > m_c >> m_q, m_{q^\prime}$ limit and in the
neighborhood of $w=1$, the ${\cal I}$ and ${\cal K}$ baryon factors
behave like ${\cal O}(1)$ and ${\cal O}(m_q/m_c, m_{q^\prime}/m_c)$,
respectively\footnote{This is trivial, since in this limit for the ${\cal
I}$ integral case, the $l^{\prime\prime}=0$ contribution becomes the
dominant one, while for the ${\cal K}$ factor, the
$l^{\prime\prime}=l^{\prime\prime\prime}=0$ contribution is forbidden
by the Clebsh-Gordan $(l^{\prime\prime\prime}
l^{\prime\prime}1|000)$. Thus, for this latter baryon factor and in
the mentioned limit, the leading contributions are the
$l^{\prime\prime}=1, l^{\prime\prime\prime}=0$ and
$l^{\prime\prime}=0, l^{\prime\prime\prime}=1$ ones.}.
 
\begin{acknowledgments}
This research was supported by DGI and FEDER funds, under contracts
BFM2002-03218, BFM2003-00856 and FPA2004-05616,  by the Junta de Andaluc\'\i a and
Junta de Castilla y Le\'on under contracts FQM0225 and
SA104/04, and it is part of the EU
integrated infrastructure initiative
Hadron Physics Project under contract number
RII3-CT-2004-506078.  C. Albertus wishes to acknowledge a grant 
 from Junta de Andaluc\'\i a.
\end{acknowledgments}

\end{document}